\documentclass[11pt,onecolumn,twoside,draftclsnofoot]{IEEEtran}

\usepackage{amsfonts,amsmath,amssymb,graphicx}
\usepackage{setspace}
\usepackage{graphicx}
\usepackage{subfigure}
\usepackage[font=bf,skip=4pt]{caption}
\usepackage{algorithmicx}
\usepackage{algorithm}
\usepackage{cite}
\usepackage{url}
\usepackage{algorithm}
\usepackage[noend]{algpseudocode}
\usepackage{color}
\usepackage{multicol}
\usepackage{multirow}
\usepackage{xspace}
\usepackage{comment}
\usepackage{color}
\usepackage{textcomp}
\usepackage[numbers]{natbib}

\allowdisplaybreaks[1]
\ifCLASSINFOpdf
\else
\fi
\newtheorem{theorem}{Theorem}

\newcommand{\xref}[1]{\S\ref{#1}}

\usepackage{ifthen}
\newboolean{showcomments}
\setboolean{showcomments}{true}
\newcommand{\adam}[1]{\ifthenelse{\boolean{showcomments}}
{ \textcolor{cyan}{(Adam says:  #1)}}{}}
\newcommand{\xiaoqi}[1]{\ifthenelse{\boolean{showcomments}}
{ \textcolor{cyan}{(Xiaoqi says:  #1)}}{}}
\newcommand{\juba}[1]{\ifthenelse{\boolean{showcomments}}
{ \textcolor{cyan}{(Juba says:  #1)}}{}}
\newcommand{\palma}[1]{\ifthenelse{\boolean{showcomments}}
{ \textcolor{cyan}{(Palma says:  #1)}}{}}
\newcommand{\addcites}[0]{\ifthenelse{\boolean{showcomments}}
{ \textcolor{blue}{(add citation(s))}}{}}
\newcommand{\addref}[0]{\ifthenelse{\boolean{showcomments}}
{ \textcolor{green}{(add ref)}}{}}
\newcommand{\todo}[1]{\ifthenelse{\boolean{showcomments}}
{ \textcolor{red}{(To do:  #1)}}{}}

\DeclareMathOperator*{\argmin}{arg\,min}

\newcommand{\name}{{\small\sf Datum}~}
\newcommand{\nameNospace}{{\small\sf Datum}}

\newcommand{\nameSubSec}{\large\sf Datum~}

\begin{document}
%
\title{Joint Data Purchasing and Data
Placement in a Geo-Distributed Data Market
}

\author{\IEEEauthorblockN{Xiaoqi Ren, 
Palma London, 
Juba Ziani,
Adam Wierman
}
\vspace{0.1in}
\\
\IEEEauthorblockA{
California Institute of Technology\\
\{xren, plondon, jziani, adamw\}@caltech.edu}
}

\maketitle

\begin{abstract}
This paper studies two design tasks faced by a geo-distributed cloud data market: which data to purchase (data purchasing) and where to place/replicate the data for delivery (data placement). We show that the joint problem of data purchasing and data placement within a cloud data market can be viewed as a facility location problem, and is thus NP-hard. However, we give a provably optimal algorithm for the case of a data market made up of a single data center, and then generalize the structure from the single data center setting in order to develop a near-optimal, polynomial-time algorithm for a geo-distributed data market. The resulting design, \nameNospace, decomposes the joint purchasing and placement problem into two subproblems, one for data purchasing and one for data placement, using a transformation of the underlying bandwidth costs.  We show, via a case study, that \name is near-optimal (within 1.6\%) in practical settings.

\end{abstract}

\section{Introduction}\label{sec:introduction}
Ten years ago computing infrastructure was a \emph{commodity}~--~the key bottleneck for new tech startups was the cost of acquiring and scaling computational power as they grew.  Now, computing power and memory are \emph{services} that can be cheaply subscribed to and scaled as needed via cloud providers like Amazon EC2, Microsoft Azure, etc.

We are beginning the same transition with respect to \emph{data}. Data is broadly being gathered, bought, and sold in various marketplaces. However, it is still a commodity, often obtained through offline negotiations between providers and companies.  Thus,  acquiring data is one of the key bottlenecks for new tech startups nowadays.

This is beginning to change with the emergence of \emph{cloud data markets}, which offer a single, logically centralized point for buying and selling data. Multiple data markets have recently emerged in the cloud, e.g., Microsoft Azure DataMarket~\cite{Azure}, Factual~\cite{Factual}, InfoChimps~\cite{Infochimps}, Xignite~\cite{Xignite}, IUPHAR~\cite{IUPHAR}, etc. These marketplaces enable data providers to sell and upload data and clients to request data from multiple providers (often for a fee) through a unified query interface.  They provide a variety of services: (i) \emph{aggregation} of data from multiple sources, (ii) \emph{cleaning} of data to ensure quality across sources, (iii) \emph{ease of use}, through a unified API, and (iv) \emph{low-latency delivery} through a geographically distributed content distribution network.

Given the recent emergence of data markets, there are widely differing designs in the marketplace today, especially with respect to pricing.  For example, The Azure DataMarket~\cite{Azure} sets prices with a subscription model that allows a maximum number of queries (API calls) per month and limits the size of records that can be returned for a single query.  Other data markets, e.g., Infochimps~\cite{Infochimps}, allow payments per query or per data set.  In nearly all cases, the data provider and the data market operator each then get a share of the fees paid by the clients, though how this share is arrived at can differ dramatically across data markets.  The task of pricing is made even more challenging when one considers that clients may be interested in data with differing levels of precision/quality and privacy may be a concern.

Not surprisingly, the design of pricing (both on the client side and the data provider side) has received significant attention in recent years, including pricing of per-query access~\cite{koutris2012query, koutris2013toward} and pricing of private data~\cite{fleischer2012approximately, li2014theory}.

In contrast, the focus of this paper is \emph{not} on the design of pricing strategies for data markets.  Instead, \textbf{we focus on the engineering side of the design of a data market}, which has been ignored to this point.  Supposing that prices are given, there are important challenges that remain for the operation of a data market.  Specifically, two crucial challenges relate to \emph{data purchasing} and \emph{data placement}.


\vspace{2 pt}
\noindent\textbf{Data purchasing:} Given prices and contracts offered by data providers, which providers should a data market purchase from to satisfy a set of client queries with minimal cost?

\vspace{2 pt}\noindent\textbf{Data placement:} How should purchased data be stored and replicated throughout a geo-distributed data market in order to minimize bandwidth and latency costs? And which clients should be served from which replicas given the locations and data requirements of the clients?
\vspace{2 pt}

Clearly, these two challenges are highly related: data placement decisions depend on which data is purchased from where, so the bandwidth and latency costs incurred because of data placement must be balanced against the purchasing costs.  Concretely, less expensive data that results in larger bandwidth and latency costs is not desirable.

\textbf{The goal of this paper is to present a design for a geo-distributed data market that jointly optimizes data purchasing and data placement costs. } The combination of data purchasing and data placement decisions makes the task of operating a geo-distributed data market more complex than the task of operating a geo-distributed data analytics system, which has received considerable attention in recent years e.g.,~\cite{pu2015low,vulimiri2015wanalytics,vulimiri2015global}. Geo-analytics systems minimize the cost (in terms of latency and bandwidth) of moving the data needed to answer client queries, replacing the traditional operation mode where data from multiple data centers was moved to a central data center for processing queries.  However, crucially, such systems do not consider the cost of obtaining the data (including purchasing and transferring) from data providers.

Thus, the design of a geo-distributed data market necessitates integrating data purchasing decisions into a geo-distributed data analytics system.  To that end, our design builds on the model used in~\cite{vulimiri2015global} by adding data providers that offer a menu of data quality levels for differing fees.  The data placement/replication problem in~\cite{vulimiri2015global} is already an integer linear program (ILP), and so it is no surprise that the addition of data providers makes the task of jointly optimizing data purchasing and data placement NP-hard (see Theorem~\ref{T:REDUCTION}).

Consequently, we focus on identifying structure in the problem that can allow for a practical and near-optimal system design.  To that end, we show that the task of jointly optimizing data purchasing and data placement is equivalent to the uncapacitated facility location problem (UFLP)~\cite{Krarup83}.  However, while constant-factor polynomial running time approximation algorithms are known for the \emph{metric} uncapacitated facility location problem (see \cite{CGTS99, GK99, JKV01}), our problem is a \emph{non-metric} facility location problem, and the best known polynomial running time algorithms achieve a $O(\log{C})$ approximation via 
the greedy algorithm in~\cite{Hochbaum82} or the randomized rounding algorithm in~\cite{Vazirani2001}, where $C$ is the number of clients.  Note that without any additional information on the costs, this approximation ratio is the smallest achievable for the non-metric uncapacitated facility location  unless NP has slightly superpolynomial time algorithms~\cite{Feige98}. 
While this is the best theoretical guarantee possible in the worst-case, some promising heuristics have been proposed for the non-metric case, e.g.,~\cite{Erlenkotter78, Beasley93, SF99, Korkel89, Tuzun99, Ghosh03}.

Though the task of jointly optimizing data purchasing and data placement is computationally hard in the worst case, in practical settings there is structure that can be exploited. In particular, we provide an algorithm with polynomial running time that gives an exact solution in the case of a data market with a single data center (\xref{subsec:singledc}). Then, using this structure, we generalize to the case of a geo-distributed data cloud and provide an algorithm, named~\name (\xref{SUBSEC:HEURISTIC}) that is near optimal in practical settings.

\name first optimizes data purchasing as if the data market was made up of a single data center (given carefully designed ``transformed'' costs) and then, given the data purchasing decisions, optimizes data placement/replication.  The ``transformed'' costs are designed to allow an architectural decomposition of the joint problem into subproblems that manage data purchasing (external operations of the data market) and data placement (internal operations of the data market).  This decomposition is of crucial operational importance because it means that internal placement and routing decisions can proceed without factoring in data purchasing costs, mimicking operational structures of geo-distributed analytics systems today.

We provide a case study in \xref{sec:case_study} which highlights that \name is near-optimal (within $1.6\%$) in practical settings.  Further, the performance of \name improves upon approaches that neglect data purchasing decisions by $> 45\%$. 

To summarize, this paper makes the following main contributions:
\begin{enumerate}
\item We initiate the study of jointly optimizing data purchasing and data placement decisions in geo-distributed data markets.
\item We prove that the task of jointly optimizing data purchasing and data placement decisions is NP-hard and can be equivalently viewed as a facility location problem.
\item We provide an exact algorithm with polynomial running time for the case of a data market with a single data center.
\item We provide an algorithm, \nameNospace, for jointly optimizing data purchasing and data placement in a geo-distributed data market that is within $1.6\%$ of optimal in practical settings and improves by $>45\%$ over designs that neglect data purchasing costs. 
Importantly, \name decomposes into subproblems that manage data purchasing and data placement decisions separately.
\end{enumerate}

\section{Opportunities and challenges}
\label{sec:opps}

Data is now a traded \emph{commodity}.  It is being bought and sold every day, but most of these transactions still happen offline through direct negotiations for bulk purchases.  This is beginning to change with the emergence of cloud data markets such as Microsoft Azure DataMarket in~\cite{Azure}, Factual~\cite{Factual}, InfoChimps~\cite{Infochimps}, Xignite~\cite{Xignite}.  As cloud data markets become more prominent, data will become a \emph{service} that can be acquired and scaled seamlessly, on demand, similarly to computing resources available today in the cloud.

\subsection{The potential of data markets}

The emergence of cloud data markets has the potential to be a significant disruptor for the tech industry, and beyond.  Today, since computing resources can be easily obtained and scaled through cloud services, data acquisition has become the bottleneck for new tech startups.

For example, consider an emerging potential competitor for Yelp.  The biggest development challenge is not algorithmic or computational. Instead, it is obtaining and managing high quality data at scale. The existence of a data market, e.g., Azure DataMarket, with detailed local information about restraints, attractions, etc., would eliminate this bottleneck entirely. In fact, data markets such as Factual~\cite{Factual} are emerging to target exactly this need.

Another example highlighted in~\cite{koutris2012demonstration, koutris2013toward} is language translation.  Emerging data markets such as the Azure DataMarket and Infochimps sell access to data on word translation, word frequency, etc. across languages.  This access is a crucial tool for easing the transition tech startups face when moving into different cultural markets.

A final example considers computer vision.  When tech startups need to develop computer vision tools in house, a significant bottleneck (in terms of time and cost) is obtaining labeled images with which to train new algorithms.  Emerging data markets have the potential to eliminate this bottleneck too.  For example, the emerging Visipedia project~\cite{Visipedia} (while free for now) provides an example of the potential of such a data market.

Thus, like in the case of cloud computing, ease of access and scaling, combined with the cost efficiency that comes with size, implies that cloud data markets have the potential to eliminate one of the major bottlenecks for tech startups today -- data acquisition.

\subsection{Operational challenges for data markets}

The task of designing a cloud data market is complex, and requires balancing economic and engineering issues.  It must carefully consider purchasing and pricing decisions in its interactions with both data providers and clients and minimize its operational cost, e.g., from bandwidth.  We discuss both the economic and engineering design challenges below, though this paper focuses only on the engineering challenges.

\subsubsection{Pricing}\label{subsec:price}


While there is a large body of literature on selling physical goods, the problem of pricing digital goods, such as data, is very different.  Producing physical goods usually has a moderate fixed cost, for example, for buying the space and production machines needed, but this cost is partly recoverable: it is possible, if the company cannot manage to sell its product, to resell the machinery and buildings they have been using. However, the cost of producing and acquiring data is high and irrecoverable: if the data turns out to be worthless and nobody wants it, then the whole procedure is wasted. Another major difference comes from the fact that variable costs for data are low: once it has been produced, data can be cheaply copied and replicated. 

These differences lead to ``versioning'' as the most typical approach for selling digital goods~\cite{balazinska2013discussion}.  Versioning refers to selling different versions of the same digital good at different prices in order to target different types of buyers. This pricing model is common in the tech industry, e.g., companies like Dropbox sell digital space at different prices depending on how much space customers need and streaming websites such as Amazon often charge differently for streaming movies at different quality levels.

In the context of data markets, versioning is also common.  For example, in Infochimps and the Azure DataMarket data consumers may pay a monthly subscription fee that varies according to the maximum number of queries they are allowed to run.  Additionally, when charging per query, proposals have suggested it is desirable to charge based on the complexity of the query, e.g., ~\cite{balazinska2011data,balazinska2013discussion}.  Another form of versioning that has been proposed in data markets deals with privacy --  data with more personal information should be charged more, e.g.,~\cite{li2014theory,cummings2015accuracy}.

There is a growing literature focused on the design of pricing strategies for cloud data markets in the above, and other contexts, e.g.,~\cite{balazinska2011data, koutris2012query, balazinska2013discussion, koutris2013toward, balazinska2013discussion, li2014theory,cummings2015accuracy}.

\subsubsection{Data purchasing and data placement}\label{subsec:data_placement}

While data pricing within cloud data markets has received increasing attention, the engineering of the system itself has been ignored.  The engineering of such a geo-distributed ``data cloud'' is complex.  In particular, the system must jointly make both data purchasing decisions and data placement, replication and delivery decisions, as described in the introduction.

Even considered independently, the task of optimizing data placement/replication within a geo-distributed data analytics system is challenging.  Such systems aim to allow queries on databases that are stored across data centers, as opposed to traditional databased that are stored within a single data center.  Examples include Google Spanner~\cite{corbett2013spanner}, Mesa~\cite{gupta2014mesa}, JetStream~\cite{rabkin2014aggregation}, Geode~\cite{vulimiri2015global}, and Iridium~\cite{pu2015low}.  The aim in designing a geo-distributed data analytics system is to distribute the computation needed to answer queries across data centers; thus avoiding the need to transfer all the data to a single data center to respond to queries.  This distribution of computation is crucial for minimizing bandwidth and latency costs, but leads to considerable engineering challenges, e.g., handling replication constraints due for fault tolerance and regulatory constraints on data placement due to data privacy.  See~\cite{vulimiri2015global, pu2015low} for a longer discussion of these challenges and for examples illustrating the benefit of distributed query computation in geo-distributed data analytics systems.


Importantly, all previous work on geo-distributed analytics systems assumes that the system already owns the data.  Thus, on top of the complexity in geo-distributed analytics systems, a geo-distributed cloud data market must balance the cost of data purchasing with the impact on data placement/replication costs as well as the decisions for data delivery. For example, if clients who are interested in some data are located close to data center $A$, while the data provider is located close to data center $B$  (far from data center $A$), it may be worth it to place that data in data center $A$ rather than data center $B$. In practice, the problem is more complex since clients are usually geo-graphically distributed rather than centralized and one client may require data from  several different data providers.

Additional complexity is created by versioning the data, i.e., the fact that clients have differing quality requirements for the data requested.  For example, if some clients are interested in high quality data and others are interested in low quality data, then it may be worth it to provide high quality level data to some clients that only need low quality data (thus incurring a higher price) because of the savings in bandwidth and replication costs that result from being able to serve multiple clients with the same data.

\section{A Geo-Distributed Data Cloud}
This paper presents a design for a geo-distributed cloud data market, which we refer to as a ``data cloud." This data cloud serves as an intermediary between \emph{data providers}, which gather data and offer it for sale, and \emph{clients}, which interact with the data cloud through queries for particular subsets/qualities of data. More concretely, the data cloud purchases data from multiple data providers, aggregates it, cleans it, stores it (across multiple geographically distributed data centers), and delivers it (with low-latency) to clients in response to queries, while aiming at minimizing the \emph{operational cost} constituted of both bandwidth and data purchasing costs. 

Our design builds on and extends the contributions of recent papers -- specifically~\cite{vulimiri2015global, pu2015low} -- that have focused on building geo-distributed data analytic systems but assume the data is already owned by the system and focus solely on the interaction between a data cloud and its clients. Unfortunately, as we highlight in \xref{sec:opt_purchasing_payment}, the inclusion of data providers means that the data cloud's goal of cost minimization can be viewed as a non-metric uncapacitated facility location problem, which is NP-hard.

For reference, Figure \ref{fig:DP-DC-C} provides an overview of the interaction between these three parties as well as some basic notations. 
\subsection{Modeling Data Providers}\label{subsec:providers}

The interaction between the data cloud and data providers is a key distinction between the setting we consider and previous work on geo-distributed data analytics systems such as~\cite{pu2015low, vulimiri2015global}. We assume that each data provider offers distinct data to the data cloud, and that the data cloud is a price-taker, i.e., cannot impact the prices offered by data providers.  Thus, we can summarize the interaction of a data provider with the data cloud through an exogenous menu of data qualities and corresponding prices.

We interpret the quality of data as a general concept that can be instantiated in multiple ways. For categorical data, quality may represent the resolution of the information provided, e.g., for geographical attributes the resolution may be \{street address, zip code, city, county, state\}. For numerical data, quality could take many forms, e.g., the numerical precision, the statistical precision (e.g., the confidence of an estimator), or the level of noise added to the data.\footnote{A common suggestion for guaranteeing privacy is to add Laplace noise to data provided to data markets, see e.g., \cite{dwork2011differential, li2014theory}}

Concretely, we consider a setting where there are $P$ data providers selling different data, $p \in \mathcal{P} = \{1, 2, \ldots, P\}$.\footnote{We distinguish data providers based on data, i.e., one data provider sells multiple data is treated as multiple data providers.} Each data provider offers a set of quality levels, indexed by $l \in \mathcal{L} = \{1, 2, \ldots, L_p\}$, where $L_p$ is the number of levels that data provider $p$ offers. We use $q(l, p)$ to denote the data quality level $l$, offered by data provider $p$.  Similarly, we use $f(l, p)$ to denote the fee charged by data provider $p$ for data of quality level $l$. Importantly, the prices vary across providers $p$ since different providers have different procurement costs for different qualities and different data. 

The data purchasing contract between data providers and data cloud may have a variety of different types. For example, a data cloud may pay data provider based on usage, i.e., per query, or a data cloud may buy the data in bulk in advance. In this paper, we discuss both per-query data contracting and bulk data contracting. See~\xref{subsubsec:cost} for details. 


\subsection{Modeling Clients}\label{subsec:clients}
Clients interact with the data cloud through queries, which may require data (with varying quality levels) from multiple data providers.

Concretely, we consider a setting where there are $C$ clients, $c \in \mathcal{C} = \{1, 2, \ldots, C\}$. A client $c$ sends a \textit{query} to the data center, requesting particular data from multiple data providers.\footnote{We distinguish clients based on queries, i.e., one client sends multiple queries is treated as multiple clients.} Denote the set of data providers required by the request from client query $c$ by $G(c)$.  The client query also specifies a minimum desired quality level, $w_c(p)$, for each data provider $p$ it requests, i.e., $\forall p \in G(c)$. We assume that the client is satisfied with data at a quality level higher than or equal to the level requested.

More general models of queries are possible, e.g., by including a DAG modeling the structure of the query and query execution planning (see~\cite{vulimiri2015global} for details).  For ease of exposition, we do not include such detailed structure here, but it can be added at the expense of more complicated notation.

Depending on the situation, the client may or may not be expected to pay the data cloud for access.  If the clients are internal to the company running the data cloud, client payments are unnecessary.   However, in many situations the client is expected to pay the data cloud for access to the data.  There are many different types of payment structures that could be considered.  Broadly, these  fall into two categories: (i) subscription-based (e.g., Azure DataMarket~\cite{Azure}) or (ii) per-query-based (e.g. Infochimps~\cite{Infochimps}).

In this paper, we do not focus on (or model) the design of payment structure between the clients and the data cloud.  Instead, we focus on the operational task of minimizing the cost of the data cloud operation (i.e., bandwidth and data purchasing costs).  This focus is motivated by the fact that minimizing the operation costs improves the profit of the data cloud regardless of how clients are charged.  Interested readers can find analyses of the design of client pricing strategies in~\cite{koutris2012query, koutris2013toward, li2014theory}.

\subsection{Modeling a Geo-Distributed Data Cloud}

The role of the data cloud in this marketplace is as an aggregator and intermediary.   We model the data cloud as a geographically distributed cloud consisting of $D$  data centers, $d \in \mathcal{D} = \{1, 2, \ldots, D\}$. Each data center  aggregates data from geographically separate local data providers, and data from data providers may be (and often is) replicated across multiple data centers within the data cloud. 

Note that, even for the same data with the same quality, data transfer from the data providers to the data cloud is not a one time event due to the need of the data providers to update the data over time. We target the modeling and optimization of data cloud within a fixed time horizon, given the assumption that queries from clients are known beforehand or can be predicted accurately. This assumption is consistent with previous work~\cite{vulimiri2015global, pu2015low} and reports from other organizations~\cite{facebook2014, lee2012unified}.  Online versions of the problem are also of interest, but are not the focus of this paper.

\begin{figure}[t!]
\centering
\includegraphics[width=8cm]{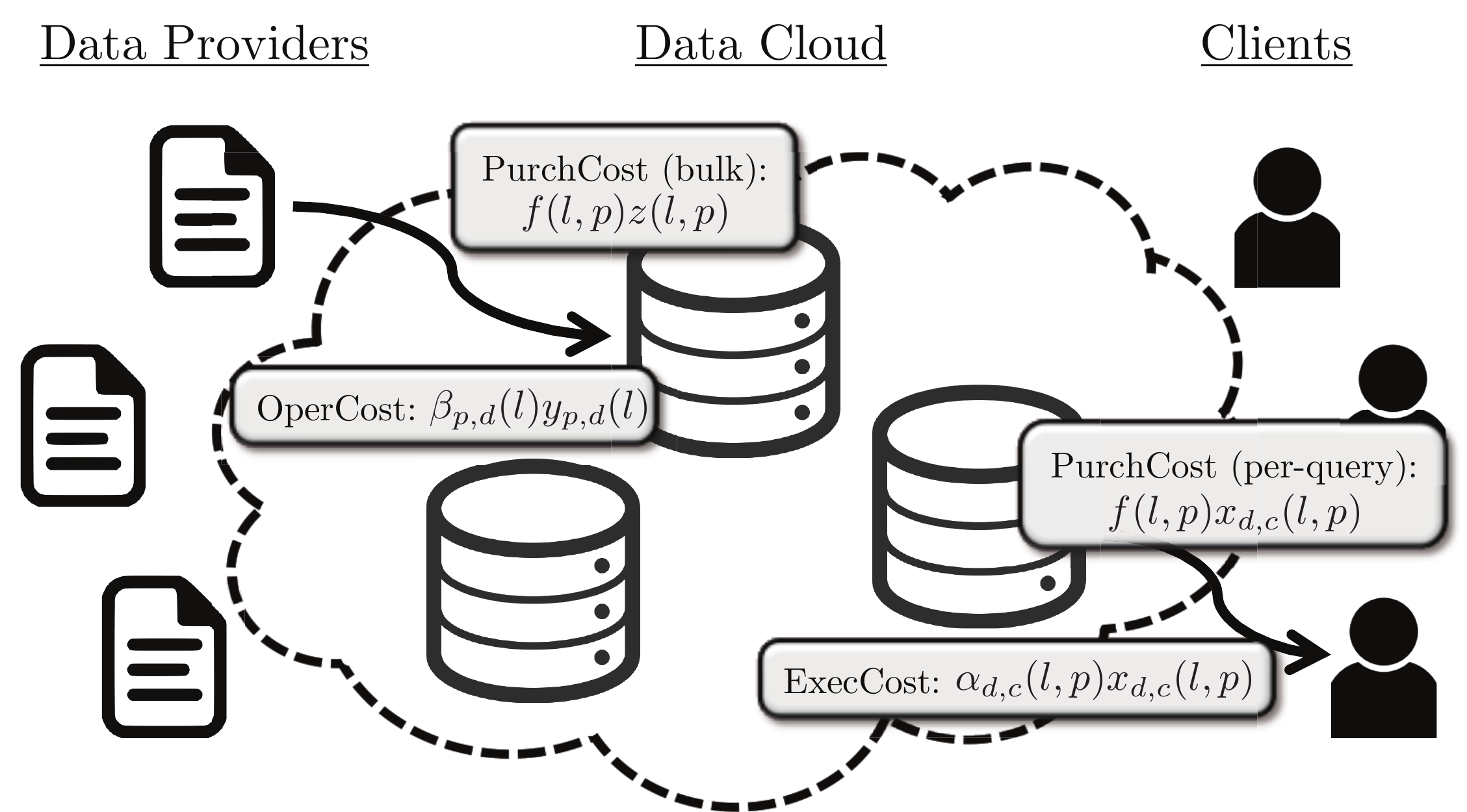}
\caption{An overview of the interaction between data providers, the data cloud, and clients. The dotted line encircling the data centers (DC) represents the geo-distributed data cloud. Data providers and clients interact only with the cloud. Data provider $p$ sends data of quality $q(l, p)$ to data center $d$, and the corresponding operation cost is $\beta_{p,d}(l) y_{p,d}(l)$. Similarly, data center $d$ sends data of quality $q(l, p)$ to client $c$, and the corresponding execution cost is $\alpha_{d,c}(l,p) x_{d,c}(l,p)$. In bulk data contracting, the corresponding purchasing cost is $f(l, p)z(l, p)$.  In per-query data contracting, the corresponding purchasing cost is $f(l, p)x_{d, c}(l, p)$.}
\label{fig:DP-DC-C}
\end{figure}


\subsubsection{Modeling costs}\label{subsubsec:cost}

Our goal is to provide a design that minimizes the operational costs of a data cloud.  These costs include both data purchasing and bandwidth costs.  In order to describe these costs, we use the following notation, which is summarized in Figure \ref{fig:DP-DC-C}.\footnote{Throughout, subscript indices refer to data transfer ``from, to'' a location, and parenthesized indices refer to data characteristics (e.g., quality, from which data provider).}\\


\noindent \hangindent=1.3cm{
$x_{d,c}(l, p)   \in \{0,1\}$: $x_{d,c}(l, p)=1$ if and only if data of quality $q(l,p)$, originating from data provider $p$, is transferred from data center $d$ to client $c$. }

\noindent \hangindent=1.3cm{
$\alpha_{d,c}(l, p)$: cost (including bandwidth and/or latency) to transfer data of quality $q(l,p)$, originating from data provider $p$, from data center $d$ to client $c$
} \\

\noindent \hangindent=1.3cm{
$y_{p,d}(l) \in \{0,1\}$: $y_{p,d}(l)=1$ if and only if data of quality $q(l,p)$ is transferred from data provider $p$ to data center $d$.}

\noindent \hangindent=1.3cm{
$\beta_{p,d}(l)$: cost (including bandwidth and/or latency) to transfer data of quality $q(l,p)$ from data provider $p$ to data center $d$. 
} \\


\noindent \hangindent=1.3cm{
$z(l, p)   \in \{0,1\}$: $z(l, p)=1$ if and only if data of quality $q(l,p)$, originating from data provider $p$, is transferred to the data cloud. }

\noindent \hangindent=1.3cm{
$f(l, p)$: purchasing cost of data with quality $q(l,p)$, originating from data provider $p$.
} \\

Given the above notations, the costs of the data cloud can be broken into three categories:
\begin{enumerate}
\item The \emph{operation cost} due to transferring data of all quality levels from data providers to data centers is
    \begin{equation}\label{eqn: oper_cost}
    \text{OperCost}=  \sum_{p=1}^{P}\sum_{l=1}^{L_p}\sum_{d=1}^{D} \beta_{p,d}(l) y_{p,d}(l).
    \end{equation}

\item The \emph{execution cost} due to transferring data of all quality levels from data centers to clients is
    \begin{equation}\label{eqn: exec_cost}
    \text{ExecCost} = \sum_{c=1}^C\sum_{p\in G(c)}\sum_{l=1}^{L_p}\sum_{d=1}^{D} \alpha_{d,c}(l, p) x_{d,c}(l, p).
    \end{equation}

\item The \emph{purchasing cost} ($\text{PurchCost}$) due to buying data from the data provider could result from a variety of differing contract styles. In this paper we consider two extreme options: \emph{per-query} and \emph{bulk} data contracting.  These are the most commonly adopted strategies for data purchasing today.

    In \emph{per-query} data contracting, the data provider charges the data cloud a fixed rate for each query that uses the data provided by the data provider.  So, if the same data is used for two different queries, then the data cloud pays the data provider twice. Given a per-query fee $f(l,p)$ for data $q(l,p)$, the total purchasing cost is
            \begin{equation}\label{eqn: purch_cost}
            \text{ PurchCost(query)} = \sum_{c=1}^C\sum_{p\in G(c)}\sum_{l=1}^{L_p}\sum_{d=1}^{D}  f(l,p) x_{d,c}(l,p).
            \end{equation}

    In \emph{bulk} data contracting, the data cloud purchases the data in bulk and then can distribute it without owing future payments to the data provider. Given a one-time fee $f(l,p)$ for data $q(l,p)$, the total purchasing cost is
            \begin{equation}\label{eqn: bulk_fee}
            \text{PurchCost(bulk)} = \sum_{p=1}^{P}  \sum_{l=1}^{L_p} f(l,p) z(l, p).
            \end{equation}

\end{enumerate}

To keep the presentation of the paper simple, we focus on the per-query data contracting model throughout the body of the paper and discuss the bulk data contracting model (which is simpler) in Appendix~\ref{app:bulk}.

\subsubsection{Cost Optimization}
Given the cost models described above, we can now represent the goal of the data cloud via the following integer linear program (ILP), where OperCost, ExecCost, and PurchCost are as described in equations~\eqref{eqn: oper_cost},~\eqref{eqn: exec_cost} and~\eqref{eqn: purch_cost}, respectively.
\begin{subequations}\label{eqn: ILP}
\begin{align}
\underset{x, y} {\text{min}} \quad &
\text{OperCost} +\text{ExecCost}+\text{PurchCost} \tag{\ref{eqn: ILP}}\\
 \text{subject to} \quad \label{ILP: x_less_than_y}
&x_{d,c}(l,p) \le y_{p,d}(l)\;\forall c, p, l, d \\
\label{ILP: purchase_decision}
&\sum_{l=1}^{L_p}\sum_{d=1}^Dx_{d,c}(l,p) = 1, \;\forall c, p \in G(c)\\
\label{ILP: quality}
&\sum_{l=1}^{L_p}\sum_{d=1}^Dx_{d,c}(l,p) q(l, p) \ge w_c(p), \;\forall c, p \in G(c)\\
&x_{d,c}(l,p) \ge 0, \forall c, p, l, d\\
&y_{p,d}(l) \ge 0, \;\forall p, l, d\\
\label{problem:1}
&x_{d,c}(l,p), y_{p,d}(l) \in \{0, 1\}, \forall c, p, l, d
\end{align}
\end{subequations}

The constraints in this formulation warrant some discussion. Constraint~\eqref{ILP: x_less_than_y} states that any data transferred to some client must already have been transferred from its data provider to the data cloud.\footnote{For bulk data contracting model, one more constraint $y_{p, d}(l) \le z(l, p), \;\forall c, l, p, d$ is required. This constraint states that any data placed in the data cloud must be purchased by the data cloud. } Constraint~\eqref{ILP: purchase_decision} ensures that each client must get the data it requested, and constraint~\eqref{ILP: quality} ensures that the minimum quality requirement of each client must be satisfied. The remaining constraints state that the decision variables are binary and nonnegative.

An important observation about the formulation above is that data purchasing/placement decisions are decoupled across data providers, i.e., the data purchasing/placement decision for data from one data provider does not impact the data purchasing/placement decision for any other data providers. Thus, we frequently drop the index $p$.

Note that there are a variety of practical issues that we have not incorporated into the formulation in~\eqref{eqn: ILP} in order to minimize notational complexity, but which can be included without affecting the results described in the following.  A first example is that a minimal level of data replication is often desired for fault tolerance and disaster recovery reasons.  This can be added to~\eqref{eqn: ILP} by additionally considering constraints of the form $\sum_{d=1}^D y_{p, d}(l) \ge k z(l, p)$, where $k$ denotes the minimum required number of copies.  Similarly, privacy concerns often lead to regulatory constraints on data movement.  As a result, regulatory restrictions may prohibit some data from being copied to certain data centers, thus constraining data placement and replication.  This can be included by adding constraints of the form $y_{p, d}(l) = 0$ to~\eqref{eqn: ILP} where $p$ and $d$ denote the corresponding data provider and data center, respectively. Finally, in some cases it is desirable to enforce SLA constraints on the latency of delivery to clients.  Such constraints can be added by including constraints of the form $\sum_{p\in G(c)}\sum_{l=1}^{L_p}\sum_{d=1}^D\alpha_{d, c}(l, p)x_{d,c}(l,p) \le r_c$, where $r_c$ denotes the SLA requirement of client $c$.

We refer the reader to~\cite{vulimiri2015global, vulimiri2015wanalytics, pu2015low} for more discussions of these additional practical constraints.  Each paper includes a subset of these factors in the design of geo-distributed data analytics systems, but does not model data purchasing decisions.

\section{Optimal data purchasing and data placement}\label{sec:opt_purchasing_payment}

Given the model of a geo-distributed data cloud described in the previous section, the design task is now to provide an algorithm for computing the optimal data purchasing and data placement/replication decisions, i.e., to solve data cloud cost minimization problem in~\eqref{eqn: ILP}.  Unfortunately, this cost minimization problem is an ILP, which are computationally difficult in general.\footnote{Note that previous work on geo-distributed data analytics where data providers and data purchasing were not considered already leads to an ILP with limited structure.  For example, \cite{vulimiri2015global} suggest only heuristic algorithms with no analytic guarantees.}

A classic NP-hard ILP is the uncapacitated facility location problem (UFLP)~\cite{Krarup83}.  In the uncapacitated facility location problem, there is a set of $I$ clients and $J$ potential facilities. Facility $j\in J$ costs $f_j$ to open and can serve clients $i\in I$ with cost $c_{i,j}$.  The task is to determine the set of facilities that serves the clients with minimal cost.

Our first result, stated below, highlights that cost minimization for a geo-distributed data cloud can be reduced to the uncapacitated facility location problem, and vice-versa.  Thus, the task of operating a data cloud can then be viewed as a facility location problem, where opening a facility parallels purchasing a specific quality level from a data provider and placing it in a particular data center in the data cloud.

\begin{theorem}\label{T:REDUCTION}
The cost minimization problem for a geo-distributed data cloud given in \eqref{eqn: ILP} is NP-hard.
\end{theorem}

The proof of Theorem~\ref{T:REDUCTION} (given in Appendix~\ref{app:reduction}) provides a reduction both to and from the uncapacitated facility location problem. Importantly, the proof of Theorem~\ref{T:REDUCTION} serves a dual purpose: it both characterizes the hardness of the data cloud cost minimization problem and highlights that algorithms for the facility location problem can be applied in this context.  Given the large literature on facility location, this is important.

More specifically, the reduction leading to Theorem~\ref{T:REDUCTION} highlights that the data cloud optimization problem is equivalent to the \emph{non-metric} uncapacitated facility location problem -- every instance of any of the two problems can be written as an instance of the other. While constant-factor polynomial running time approximation algorithms are given for the \emph{metric} uncapacitated facility location problem in~\cite{CGTS99, GK99, JKV01}, in the more general \emph{non-metric} case the best known polynomial running time algorithm achieves a $\log(C)$-approximation via a greedy algorithm with polynomial running time, where $C$ is the number of clients~\cite{Hochbaum82}. This is the best worst-case guarantee possible (unless NP has slightly superpolynomial time algorithms, as proven in~\cite{Feige98}); however some promising heuristics have been proposed for the non-metric case, e.g.,~\cite{Erlenkotter78, Beasley93, SF99, Korkel89, Tuzun99, Ghosh03}.

Nevertheless, even though our problem can, in general, be viewed as the non-metric uncapacitated facility location, it does have a structure in real-world situations that we can exploit to develop practical algorithms.

In particular, in this section we begin with the case of a data cloud made up of a single data center.  We show that, in this case, there is a structure that allows us to design an algorithm with polynomial running time that gives an exact solution (\xref{subsec:singledc}). Then, we move to the case of a data cloud made up of geo-distributed data centers and highlight how to build on the algorithm for the single data center case to provide an algorithm, \nameNospace,  for the general case (\xref{SUBSEC:HEURISTIC}).  Importantly, \name allows decomposition of the management of data purchasing (operations outside of the data cloud) and data placement (operations inside the data cloud).  This feature of \name is crucial in practice because it means that the algorithm allows a data cloud to manage internal operations without factoring in data purchasing costs, mimicking operations today.  While we do not provide analytic guarantees for \name (as expected given the reduction to/from the non-metric facility location problem), we show  that the heuristic performs well in practical settings using a case study in~\xref{sec:case_study}.

\subsection{An exact solution for a single data center}\label{subsec:singledc}



We begin our analysis by focusing on the case of a single data center, which interacts with multiple data providers and multiple clients. 
The key observation is that, if the execution costs associated with transferring different quality levels of the same data are the same, i.e., $\forall l, \alpha_c(l) = \alpha_c,$ then the execution cost becomes a constant which is independent of the data purchasing and data placement decisions as shown in~\eqref{eq: ExecCost = const.}.
\begin{equation}
\small
\label{eq: ExecCost = const.}
\text{ExecCost} =\sum_{c=1}^{C}\sum_{l=1}^{L}  \alpha_c x_c(l) = \sum_{c=1}^{C} \alpha_c\left(\sum_{l=1}^{L} x_c(l)\right) = \sum_{c=1}^{C} \alpha_c
\end{equation}


The assumption that the execution costs are the same across quality levels is natural in many cases. For example, if quality levels correspond to the level of noise added to numerical data, then the size of the data sets will be the same.  We adopt this assumption in what follows.

This assumption allows the elimination of the execution cost term from the objective. Additionally, we can simplify notation by removing the index $d$ for the data center.  Thus, in per-query data contracting, the data cloud optimization problem can be simplified to~\eqref{eqn: ILP_single}. (We discuss the case of bulk data contracting in Appendix~\ref{app:bulk}.)

\begin{subequations}\label{eqn: ILP_single}
\begin{align}
\text{minimize}\quad& \sum_{l=1}^L\beta(l)y(l) + \sum_{c=1}^C\sum_{l=1}^L f(l)x_c(l)\tag{\ref{eqn: ILP_single}}\\
\text{subject to} \quad
&x_{c}(l) \le y(l),\;\forall c, l \nonumber \\
&\sum_{l=w_c}^{L} x_{c}(l) = 1, \;\forall c \label{eqn: contraction_constraint} \\
&x_{c}(l) \ge 0, \forall c, l \nonumber\\
&y(l) \ge 0, \;\forall l \nonumber\\
&x_{c}(l), y(l) \in \{0, 1\}, \forall c, l \nonumber
\end{align}
\end{subequations}

Note that constraint~\eqref{eqn: contraction_constraint} is a contraction of~\eqref{ILP: purchase_decision} and~\eqref{ILP: quality}, and simply means that any client $c$ must be given exactly one quality level above $w_c$, the minimum required quality level.\footnote{While the two constraints are equivalent for an ILP, they lead to different feasible sets when considering its LP-relaxation; in particular, facility location algorithms based on LP-relaxations such as randomized rounding algorithms need to use the contracted version of the constraints to preserve the $O(\log C)$-approximation ratio for non-metric facility location. It is equivalent to the reformulation given in Appendix~\ref{app:reduction} and does not introduce infinite costs that may lead to numerical errors.} While this problem is still an ILP, in this case there is a structure that can be exploited to provide a polynomial time algorithm that can find an exact solution.  In particular, we prove in Appendix~\ref{app:singledc} that the solution to~\eqref{eqn: ILP_single} can be found by solving the linear program (LP) given in~\eqref{eqn: ILP_LR}.

\begin{subequations}\label{eqn: ILP_LR}
\begin{align}
\text{minimize}\quad& \sum_{l=1}^L\beta(l)y(l) + \sum_{i=1}^L\sum_{l=i}^L S_if(l)\chi_i(l)\tag{\ref{eqn: ILP_LR}}\\
\text{subject to}\quad \nonumber\\
&\chi_i(l) \le y(l),\;\forall i, l\nonumber\\
&\sum_{l=i}^{L} \chi_i(l) = 1, \;\forall i\nonumber\\
&\chi_i(l) \ge 0, \forall i, l\nonumber\\
&y(l) \ge 0, \;\forall l\nonumber
\end{align}
\end{subequations}

In~\eqref{eqn: ILP_LR}, $S_i$ is the number of clients who require a minimum quality level of $i$, and $\chi_i(l) = 1$ represents clients with minimum required quality level $i$ purchase at quality level $l$.

Note that this LP is not directly obtained by relaxing the integer constraints in~\eqref{eqn: ILP_single}, but is obtained from relaxing the integer constraints in a reformulation of~\eqref{eqn: ILP_single} described in  Appendix~\ref{app:singledc}. The theorem below provides a tractable, exact algorithm for cost minimization in a data cloud made up of a single data center. (A proof is given in Appendix~\ref{app:singledc}).
\begin{theorem} \label{T:SINGLEDC} There exists a binary optimal solution to the linear relaxiation program in~\eqref{eqn: ILP_LR} which is an optimal solution of the integer program in~\eqref{eqn: ILP_single} and can be found in polynomial time.
\end{theorem}

In summary, the following gives a polynomial time algorithm which yields the optimal solution of~\eqref{eqn: ILP_single}.
\vspace{5pt}\\
\noindent\textbf{Step 1:} Rewrite~\eqref{eqn: ILP_single} in the form given by~\eqref{eqn: CIP}.
\vspace{5pt}\\
\noindent\textbf{Step 2:} Solve the linear relaxation of~\eqref{eqn: CIP}, i.e., \eqref{eqn: ILP_LR}. If it gives an integral solution, this solution is an optimal solution of~\eqref{eqn: ILP_single}, and the algorithm finishes. Otherwise, denote the fractional solution of the previous step by $\{\chi^r(l), y^r(l)\}$ and continue to the next step.
\vspace{5pt}\\
\noindent\textbf{Step 3:}  Find $m_i\in \{i, \ldots, n\}$ such that $\sum_{l=i}^{m_i-1} y^r(l) < 1$, and $\sum_{l=i}^{m_i} y^r(l) \ge 1$.  (See Appendix~\ref{app:singledc} for the existence of $\{m_i\}$.) And express $\{\chi_i(l)\}$ as a function of $\{y(l)\}$ based on~\eqref{z_i(l), l_i(l)}. Substitute the expressions of $\{\chi_i(l)\}$ with $\{y(l)\}$ in~\eqref{eqn: ILP_LR} to obtain an instance of~\eqref{eqn: P}. Solve the linear programming problem~\eqref{eqn: P} and find an optimal solution that is also an extreme point of~\eqref{eqn: P}.\footnote{This step can be finished in polynomial time~\cite{bertsimas1997introduction}. } This yields a binary optimal solution of~\eqref{eqn: P}. Use transformation~\eqref{z_i(l), l_i(l)} to get a binary optimal solution of~\eqref{eqn: ILP_LR}, which can be reformulated as an optimal solution of~\eqref{eqn: ILP_single} from the definition of $\{\chi_i(l)\}$.

%

\subsection{The design of \nameSubSec}\label{SUBSEC:HEURISTIC}

Unlike the data cloud cost minimization problem for a single data center, the general data cloud cost minimization is NP-hard. In this section,
we build on the exact algorithm for cost minimization in a data cloud made up of a single data center (\xref{subsec:singledc}) to provide an algorithm, \nameNospace,  for cost minimization in a geo-distributed data cloud.  

The idea underlying \name is to, first, optimize data purchasing decisions as if the data market was made up of a single data center (given carefully designed ``transformed'' costs), which can be done tractably as a result of Theorem~\ref{T:SINGLEDC}.  Then, second, \name optimizes data placement/replication decisions given the data purchasing decisions.






Before presenting \nameNospace, we need to reformulate the general cost minimization ILP in~\eqref{eqn: ILP}.  Recall that~\eqref{eqn: ILP} is separable across providers, thus we can consider independent optimizations for each provider, and drop the index $p$ throughout.  Second, we denote the set of all possible subsets of data centers, e.g., $\{\{d_1\}, \{d_2\}, \ldots, \{d_1, d_2\}, \{d_1, d_3\}, \ldots\}$ by $V$.\footnote{Note that, in practice, the number of data centers is usually small, e.g., $10-20$ world-wide. Further, to avoid exponential explosion of $V$, the subsets included in $V$ can be limited to only have a constant number of data centers, where the constant is determined by the maximal number of replicas to be stored.} Further, define $\beta_{v}(l) = \sum_{d\in v} \beta_d(l)$, and $\alpha_{v, c}(l) = \min_{d \in v} \{\alpha_{d, c}(l)\}$.  Given this change, we define $y_v(l) = 1$ if and only if data with quality level $l$ is placed in (and only in) data centers $d \in v$ and $x_{v, c}(l) = 1$ if and only if data with quality level $l$ is transferred to client $c$ from some data center $d \in v$. These reformulations allow us to convert~\eqref{eqn: ILP} to~\eqref{eqn: ILP_1} as following.

%

\begin{subequations}\label{eqn: ILP_1}
\begin{align}
\text{minimize}\quad& \sum_{l=1}^L \sum_{v=1}^V \beta_v(l) y_v(l)  + \sum_{c=1}^C \sum_{l=1}^L \sum_{v=1}^V \alpha_{v, c}(l) x_{v, c}(l)\nonumber\\
& + \sum_{c=1}^C \sum_{l=1}^L \sum_{v=1}^V f(l) x_{v, c}(l) \tag{\ref{eqn: ILP_1}}\\
\text{subject to} \quad
&x_{v, c}(l) \le y_v(l),\;\forall c, l\\
&\sum_{l=w_c}^{L}\sum_{v=1}^V x_{v, c}(l) = 1, \;\forall c\label{ILP_1: copy_data}\\
&\sum_{v=1}^V y_v(l) \le 1,  
\;\forall l\label{ILP_1: new_y}\\
&\sum_{v=1}^V x_{v, c}(l) \le 1, \;\forall c, l\label{ILP_1: new_x}\\
&x_{v, c}(l) \ge 0, \forall v, c, l\\
&y_{v}(l) \ge 0, \;\forall v, l\\
&x_{v, c}(l), y_{v}(l) \in \{0, 1\}, \forall v, c, l
\end{align}
\end{subequations}



Compared to~\eqref{eqn: ILP}, the main difference is that~\eqref{eqn: ILP_1} has two extra constraints~\eqref{ILP_1: new_y} and~\eqref{ILP_1: new_x}. Constraint~\eqref{ILP_1: new_y} ensures that data can only be placed in at most one subset of data centers 
 across $V$. And  constraint~\eqref{ILP_1: new_x} follows from constraint~\eqref{ILP_1: copy_data}.
Using this reformulation \name can now be explained  in two steps.
\vspace{5pt}\\
\noindent\textbf{Step 1:}  Solve~\eqref{eqn: heu_S1} while treating the geo-distributed data cloud as a single data center. Specifically, define $Y(l) = \sum_{v=1}^V y_v(l)$ and $X_c(l) = \sum_{v=1}^V x_{v, c}(l)$.  Note that, $Y(l)$ and $X_c(l)$ are $0-1$ variables from Constaint~\eqref{ILP_1: new_y} and~\eqref{ILP_1: new_x}. Further, ignore the middle term in the objective, i.e., the ExecCost.  Finally, for each quality level $l$, consider a ``transformed'' cost $\beta^*(l)$. We discuss how to define $\beta^*(l)$ below. This leaves the ``single data center'' problem~\eqref{eqn: heu_S1}. Crucially, this formulation can be solved optimally in polynomial time using the results for the case of a data cloud made up of a single data center (\xref{subsec:singledc}).
\begin{subequations}\label{eqn: heu_S1}
\begin{align}
\text{minimize}\quad&\sum_{l=1}^L \beta^*(l) Y(l) + \sum_{c=1}^C \sum_{l=1}^L  f(l) X_{ c}(l)\tag{\ref{eqn: heu_S1}}\\
\text{subject to} \quad
&X_{c}(l) \le Y(l),\;\forall c, l\nonumber\\
&\sum_{l=w_c}^{L}X_{c}(l) = 1, \;\forall c\nonumber\\
&X_{c}(l) \ge 0, \forall c, l\nonumber\\
&Y(l) \ge 0, \;\forall l\nonumber\\
&X_{c}(l), Y(l) \in \{0, 1\}, \forall c, l\nonumber
\end{align}
\end{subequations}


The remaining issue is to define $\beta^*(l)$.  Note that the reason for using transformed costs $\beta^*(l)$ instead of $\beta_v(l)$ is that $\beta_v(l)$ cannot be known precisely without also optimizing the data placement.  Thus, in defining $\beta^*(l)$ we need to anticipate the execution costs that result from data placement and replication given the purchase of data with quality level $l$.  This anticipation then allows a decomposition of data purchasing and data placement decisions. Note that the only inaccuracy in the heuristic comes from the mismatch between $\beta^*(l)$ and $\min\{\beta_v(l) + \sum_{c\in C^*(l)} \alpha_{v, c}(l)\}$ where $C^*(l)$ is the set of customers who buy at quality level $l$ in an optimal solution -- if these match for the minimizer of~\eqref{eqn: ILP} then the heuristic is exact. Indeed, in order to minimize the cost of locating quality levels to data centers, and allocating clients to data centers and quality levels, the set of data centers $v$ where an optimal solution chooses to put quality level $l$ has to minimize the cost of data transfer in the set $v$ and allocating all clients who get data at quality level $l$, i.e. $C^*(l)$, to this set of data centers $v$.

Many choices are possible for the transformed costs $\beta^*(l)$.  A conservative choice is $\beta^*(l)=\min\limits_v{\beta_v(l)}$, which results in a solution (with Step 2) whose $\text{OperCost} + \text{PurchCost}$ is a lower bound to the corresponding costs in the optimal solution of~\eqref{eqn: ILP}.\footnote{However the $\text{ExecCost}$ cannot be bounded, thus we cannot obtain a bound for the total cost. The proof of this is simple and is not included in the paper due to space limit.} However, it is natural to think that more aggressive estimates may be valuable.  To evaluate this, we have performed experiments in the setting of the case study (see \xref{sec:case_study}) using the following parametric form  {\small$\beta^*(l)$ = $\min\limits_v\{\beta_v(l) + \mu_1 \sum\limits_{l' \le l}\sum\limits_{w_c = l'}\alpha_{v,c}(l') e^{-\mu_2(l-l')}\}$}, where $\mu_1$ and $\mu_2$ are parameters.  This form generalizes the conservative choice by providing a weighting of $\alpha_{v,c}(l')$ based on the ``distance'' of the quality deviation between $l'$ and the target quality level $l$. The idea behind this is that a client is more likely to be served  data with quality level close to the requested minimum quality level of the client.  Here we use the exponential decay term $e^{-\mu_2(l-l')}$ to capture the possibility of serving the data with quality level $l$ to a client with minimum quality level $l' \le l$.   
 Interestingly, in the setting of our case study, the best design is $\mu_1=\mu_2=0$, i.e., the conservative estimate $\beta^*(l)=\min\limits_v{\beta_v(l)}$, and so we adopt this $\beta^*(l)$ in \nameNospace.
\vspace{5pt}\\
\noindent\textbf{Step 2:}  At the completion of Step 1 the solution $(X, Y)$ to~\eqref{eqn: heu_S1} determines which quality levels should be purchased and which quality level should be delivered to each client. What remains is to determine data placement and data replication levels.  To accomplish this, we substitute $(X, Y)$ into~\eqref{eqn: ILP_1}, which yields~\eqref{eqn: heu_S2}.

\begin{subequations}\label{eqn: heu_S2}
\begin{align}
\text{minimize}\quad& \sum_{l=1}^L \sum_{v=1}^V \beta_v(l) y_v(l)  + \sum_{c=1}^C \sum_{l=1}^L \sum_{v=1}^V \alpha_{v, c}(l) x_{v, c}(l)\nonumber\\
& + \sum_{c=1}^C \sum_{l=1}^L \sum_{v=1}^V f(l) x_{v, c}(l) \tag{\ref{eqn: heu_S2}}\\
\text{subject to} \quad &x_{v, c}(l) \le y_v(l),\;\forall c, l\label{heu_S2: x_less_than_y}\\
&\sum_{l=w_c}^{L}\sum_{v=1}^V x_{v, c}(l) = 1, \;\forall c\label{heu_S2: x_equal_1}\\
&\sum_{v=1}^V y_v(l) = Y(l)\\
&\sum_{v=1}^V x_{v, c}(l) = X_v(l)\\
&x_{v, c}(l) \ge 0, \forall v, c, l\\
&y_{v}(l) \ge 0, \;\forall v, l\\
&x_{v, c}(l), y_{v}(l) \in \{0, 1\}, \forall v, c, l
\end{align}
\end{subequations}

The key observation is that this is no longer a computationally hard ILP.  In fact, the inclusion of $(X, Y)$ means that it can be solved in closed form.

Let $C(l)$ denote the set of clients that purchase data with quality level $l$, i.e., $C(l) = \{c: X_c(l) = 1\}$. Then~\eqref{eqn: closed_form} gives the optimal solution of~\eqref{eqn: heu_S2}. (A proof is given in Appendix~\ref{app:heu}.)
%


%
\begin{subequations}
\label{eqn: closed_form}
\begin{flalign}
y_v(l) &=
\begin{cases}
1, &\text{if}~Y(l) = 1~\text{and}~\\
   & v = \argmin\{\beta_v(l) + \sum_{c\in C(l)} \alpha_{v, c}(l)\},\\
0, &\text{otherwise.}
\end{cases} \\
x_{v, c}(l) &=
\begin{cases}
y_v(l), &\text{if}~c \in C(l),\\
0, &\text{otherwise.}
\end{cases}
\end{flalign}
\end{subequations}

\section{Case Study}\label{sec:case_study}

We now illustrate the performance of \name using a case study of a geo-distributed data cloud running in North America.  While the setting we use is synthetic, we attempt to faithfully model realistic geography for data centers in the data cloud, data providers, and clients.  Our focus is on quantifying the overall cost (including data purchasing and bandwidth/latency costs) of \name compared to two existing designs for geo-distributed data analytics systems and the optimal. To summarize, the highlights of our analysis are
\begin{enumerate}
\item \name provides consistently lower cost ($>$ 45\% lower) than 
existing designs for geo-distributed data analytics systems. 
\item \name achieves near optimal total cost (within 1.6\%) of optimal.
\item \name achieves reduction in total cost by significantly lowering purchasing costs without sacrificing bandwidth/latency costs, which stay typically within 20-25\% of the minimal bandwidth/latency costs necessary for delivery of the data to clients.
\end{enumerate}

\subsection{Experimental setup}
The following outlines the setting in which we demonstrate the empirical performance of \nameNospace.

\begin{figure*}[!t]
\centering
\subfigure[Total Cost]
{\includegraphics[height=5.3 cm]{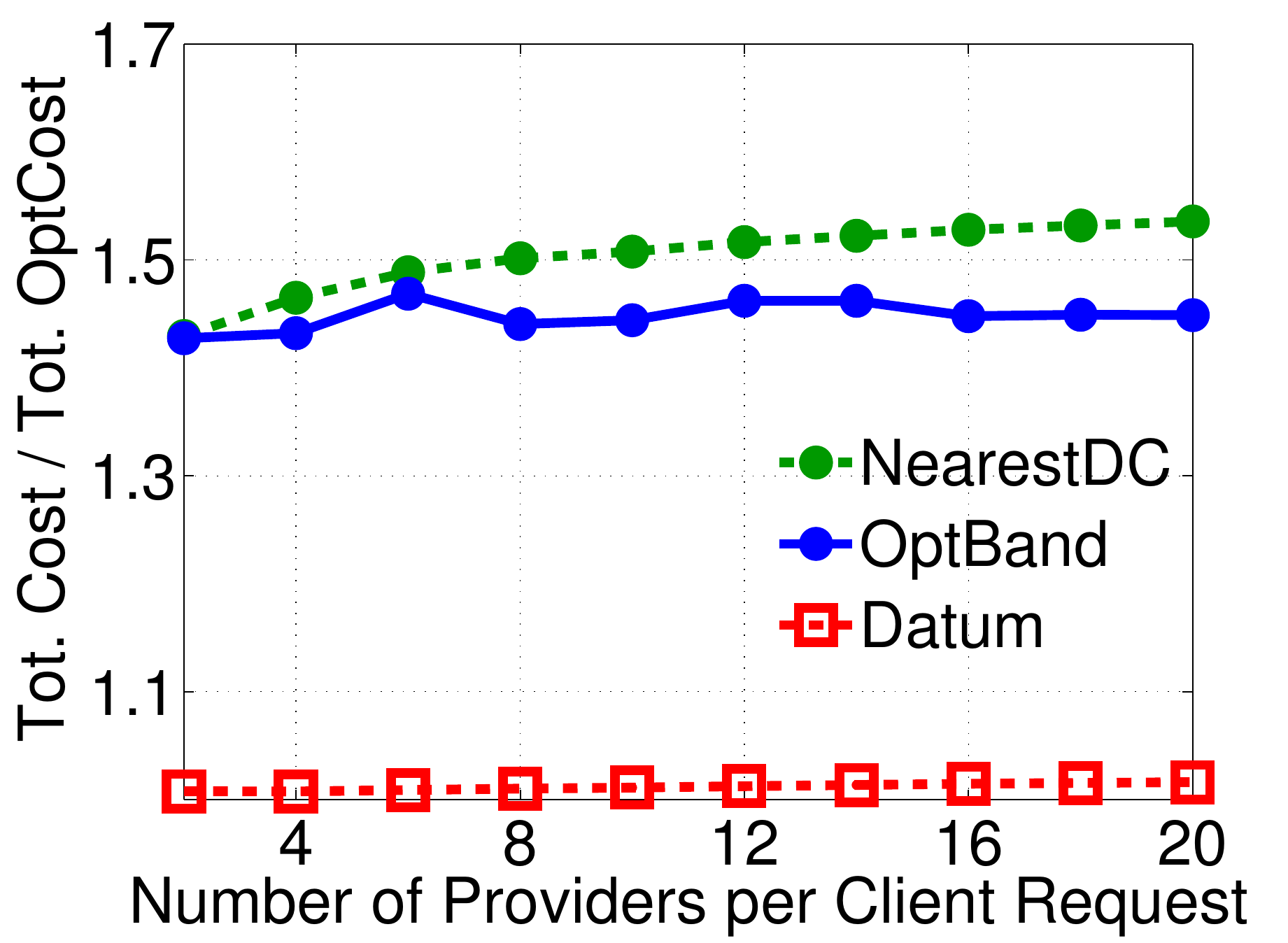}}
\hfil
\subfigure[Bandwidth Cost]
{\includegraphics[height=5.3 cm]{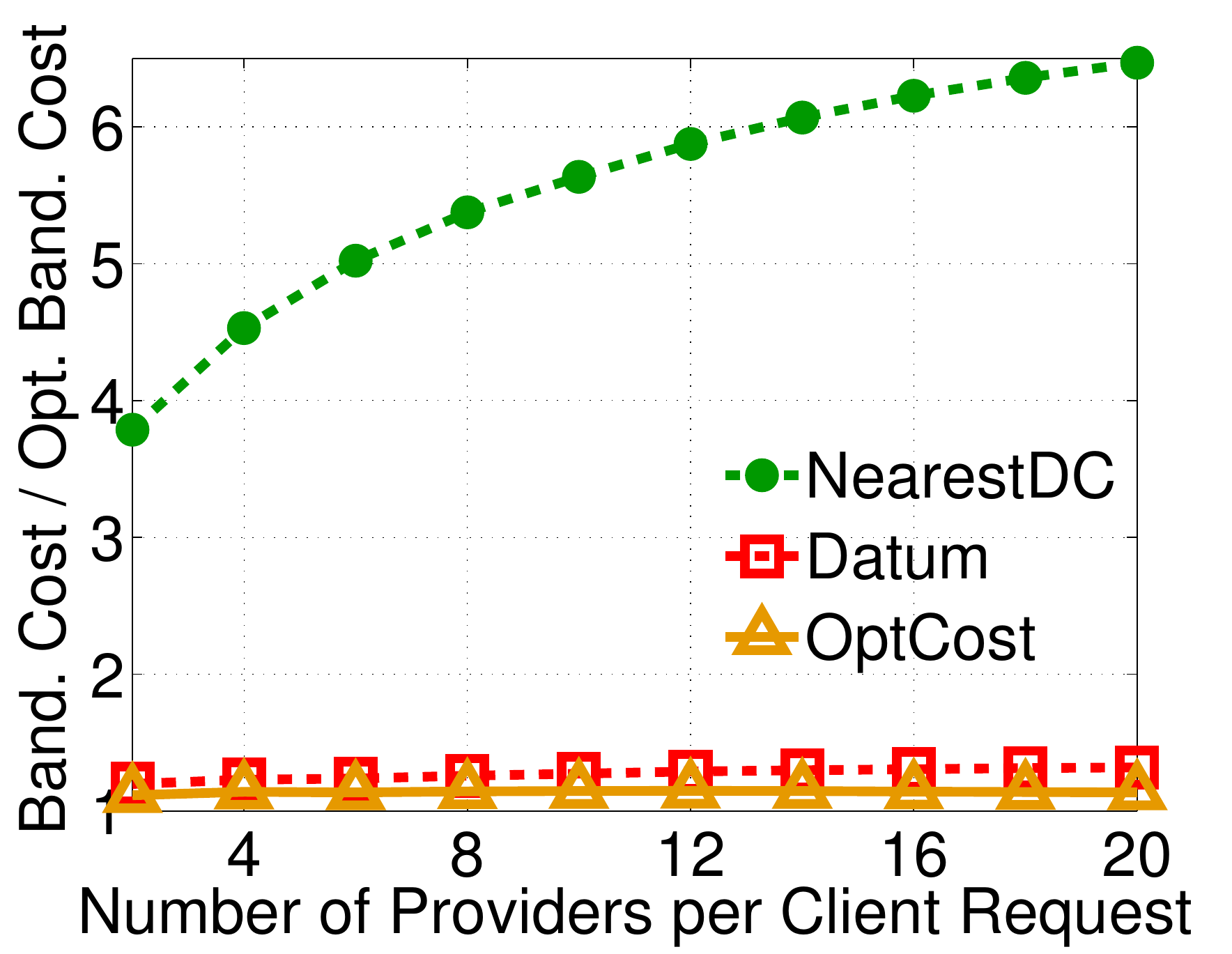}}
\caption{Illustration of the near-optimality of \name as a function of the complexity of client requests (i.e., the average number of providers data must be procured from in order to complete a client request).}
\label{fig:complexity-plots}
\end{figure*}    

\noindent\text{\bf{Geo-distributed data cloud.}}  We consider a geographically distributed data cloud with 10 data centers located in California, Washington, Oregon, Illinois, Georgia, Virginia, Texas, Florida, North Carolina, and South Carolina.  The locations of the data centers in our experiments mimic those in \cite{Datacenterknowledge} and include the locations of Google's data centers in the United States.\vspace{2pt}

\noindent\text{\bf{Clients.}}
Client locations are picked randomly among US cities, weighted proportionally to city populations. Each client requests data from a subset of data providers, chosen \emph{i.i.d.} from a Uniform distribution. Unless otherwise specified, the average number of providers per client request is $P/2$. The quality level requested from each chosen provider follows a Zipf distribution with mean $L_p/2$ and shape parameter 30. $P$ and $L_p$ are defined as in~\xref{subsec:providers} and~\xref{subsec:clients}. We choose a Zipf distribution motivated by the fact that popularity typically follows a heavy-tailed distribution \cite{newman2005power}.  Results are averaged over 20 random instances. We observe that the results of the 20 instances for the same plot are very close (within $5\%$), and thus do not show the confidence intervals on the plots.  \vspace{2pt}

\noindent\text{\bf{Data providers.}}  We consider 20 data providers.  We place data providers in the second and third largest cities within a state containing a data center.  This ensures that the data providers are near by, but not right on top of, data center and client locations. \vspace{2pt}

\noindent\text{\bf{Operation and execution costs.}}
To set operation and execution costs, we compute the geographical distances between data centers, clients and providers. The operation and execution costs are proportional to the geographical distances, such that the costs are effectively one dollar per gigameter. This captures both the form of bandwidth costs adopted in \cite{vulimiri2015wanalytics} and the form of latency costs adopted in \cite{pu2015low}.\vspace{2pt}

\noindent\text{\bf{Data purchasing costs.}} 
The per-query purchasing costs are drawn \emph{i.i.d.} from a Pareto distribution with mean 10 and shape parameter 2 unless otherwise specified.  We choose a Pareto distribution motivated by the fact that incomes and prices often follow heavy-tailed distributions \cite{newman2005power}. Results were averaged over 20 random instances. To study the sensitivity of \name to the relative size of purchasing and bandwidth costs, we vary the ratio of them between $(0.01,100)$. 

\vspace{2pt}\noindent\text{\bf{Baselines.}}
We compare the performance of \name to the following baselines.
\begin{itemize}
\item \emph{OptCost} computes the optimal solution to the data cloud cost minimization problem by solving the integer linear programming~\eqref{eqn: ILP}.  Note that this requires solving an NP-hard problem, and so is not feasible in practice.  We include it in order to benchmark the performance of \nameNospace. 
\item \emph{OptBand} computes the optimal solution to the bandwidth cost minimization problem. It is obtained by minimizing only the operation cost and execution cost in the objective of~\eqref{eqn: ILP}. Bandwidth cost minimization is commonly considered as a primary goal for cost minimization in geo-distributed data analytics systems~\cite{vulimiri2015global}.  Due to computational complexity, heuristics are usually applied to minimize the bandwidth cost. Here, instead of implementing a heuristic algorithms, we optimistically use OptBand in order to lower bound the achievable performance. Note that this also requires solving an NP-hard problem and thus is not feasible in practice.
\item \emph{NearestDC} is a greedy heuristic for the total cost minimization problem that is often applied in practice.  It serves the clients exactly what they ask for by purchasing the data and storing it at the data center closest to the data provider.
\end{itemize}

\begin{figure*}[!t]
\centering
\subfigure[]
{\includegraphics[height=5.3 cm]{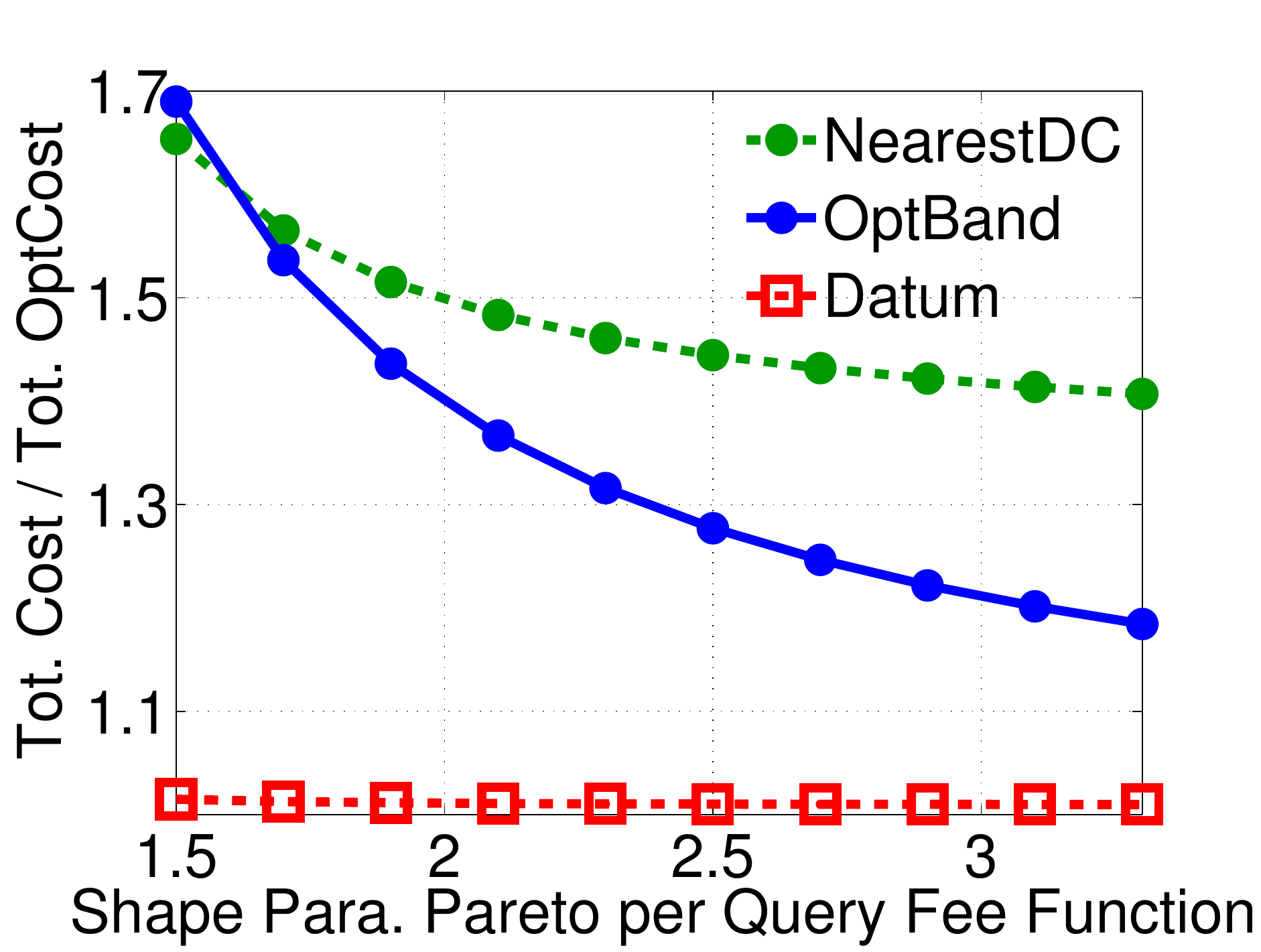}}
\hfil
\subfigure[]
{\includegraphics[height=5.3 cm]{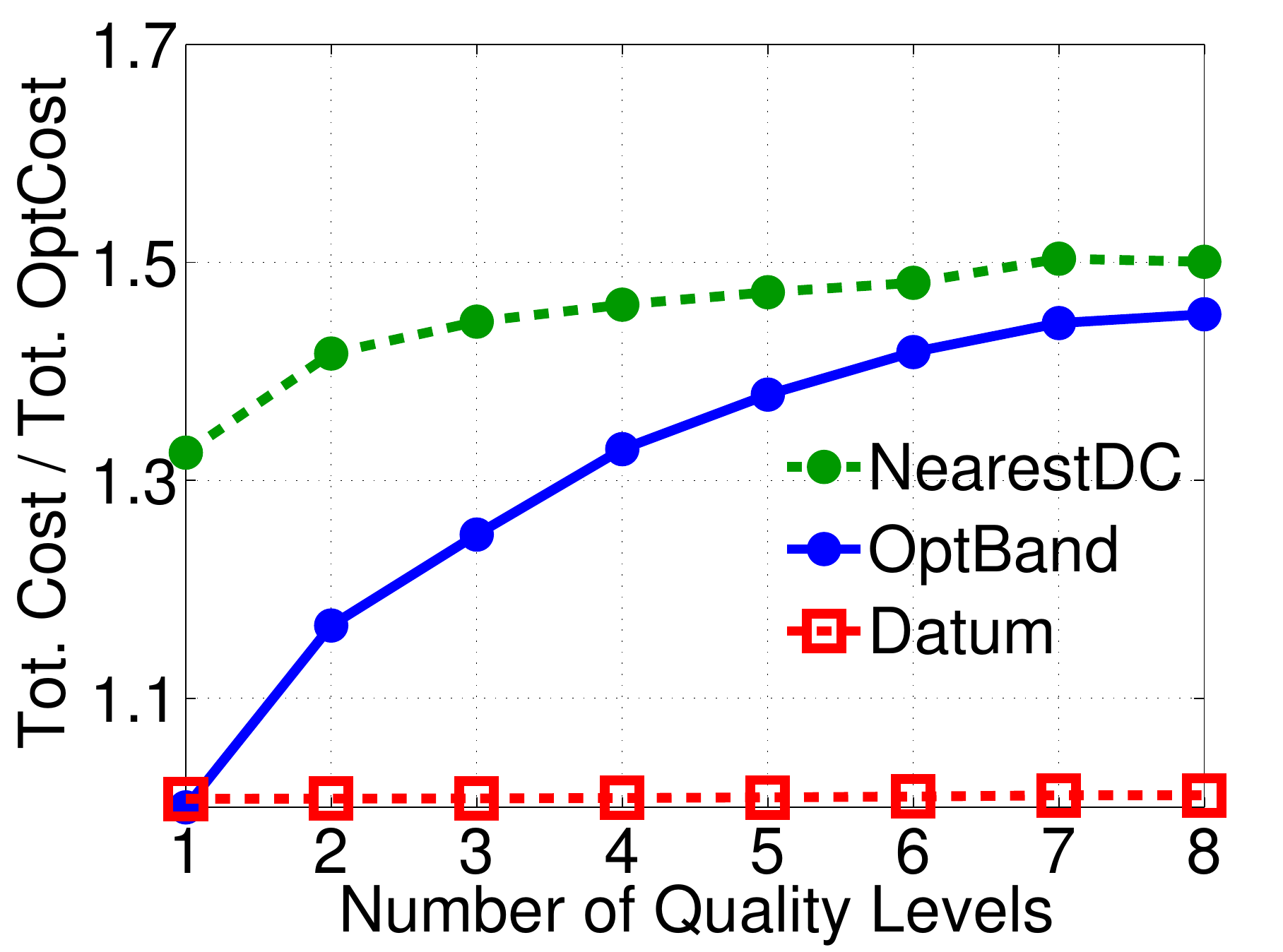}}
\caption{Illustration of \nameNospace's sensitivity to query parameters. (a) varies the heaviness of the tail in the distribution of purchasing fees. (b) varies the number of quality levels available. Note that Figure \ref{fig:complexity-plots} sets the shape parameter of the Pareto governing purchasing fees to 2 and includes 8 quality levels.}
\label{fig:para-plots}
\end{figure*}    

\subsection{Experimental results}

\noindent\textbf{Quantifying cost reductions from \nameNospace.} Figure~\ref{fig:complexity-plots}(a) illustrates the costs savings \name provides. 
Across levels of query complexity (number of providers involved), \name consistently provides $>45\%$ savings over OptBand and $>51\%$ savings compared to NearestDC.   Further, \name is within $1.6\%$ of the optimal cost in all these cases.  The improvement of \name compared to OptBand comes as a result of optimizing purchasing decisions at the expense of increased bandwidth. Importantly, Figure \ref{fig:complexity-plots}(b) shows that the extra bandwidth cost incurred is small, $20-25\%$. Thus, joint optimization of data purchasing and data placement decisions leads to significant reductions in total cost without adversely impacting bandwidth costs.

\vspace{2pt}\noindent\textbf{The form of client queries.} To understand the sensitivity of the cost reductions provided by \nameNospace, we next consider the impact of parameters related to client queries.  
Figure~\ref{fig:complexity-plots} shows that the complexity of queries has little impact on the cost reductions of \nameNospace.  Figure~\ref{fig:para-plots} studies two other parameters: the heaviness of the tail of the per-query purchasing fee and the number of quality levels offered.

Across all settings, \name is within $1.6\%$ of optimal; however both of these parameters have a considerable impact on the cost savings \name provides over our baselines.  In particular, the lighter the tail of the prices of different quality levels is, the less improvement can be achieved. This is a result of more concentration of prices across quality levels leaving less room for optimization.  Similarly, fewer quality levels provides less opportunity to optimize data purchasing decisions.  At the extreme, with only quality level available, the opportunity to optimization data purchasing goes away and OptBand and OptCost are equivalent.

\begin{figure*}[!t]
\centering
\subfigure[]
{\includegraphics[height=5.3 cm]{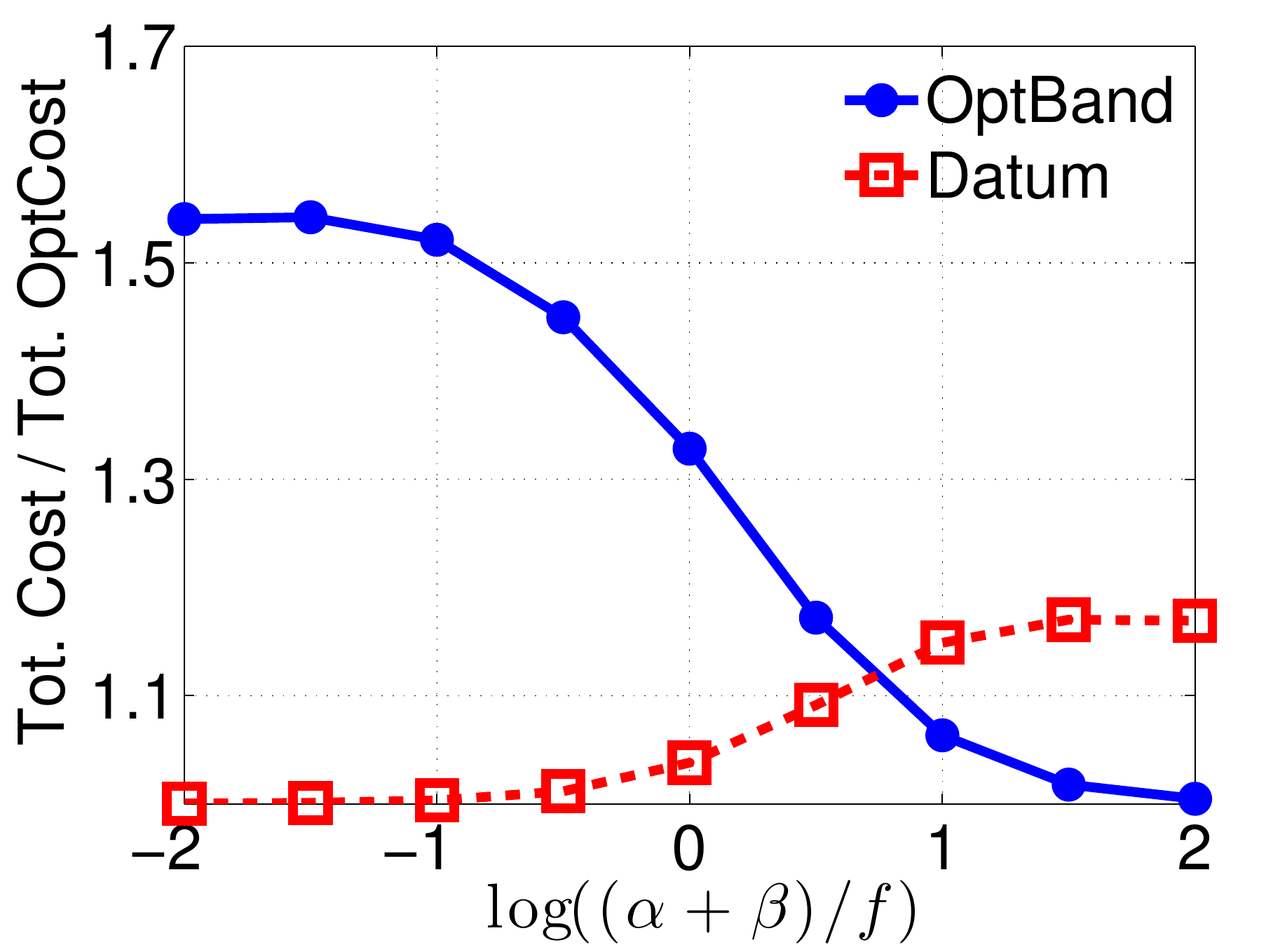}}
\hfil
\subfigure[]
{\includegraphics[height=5.3 cm]{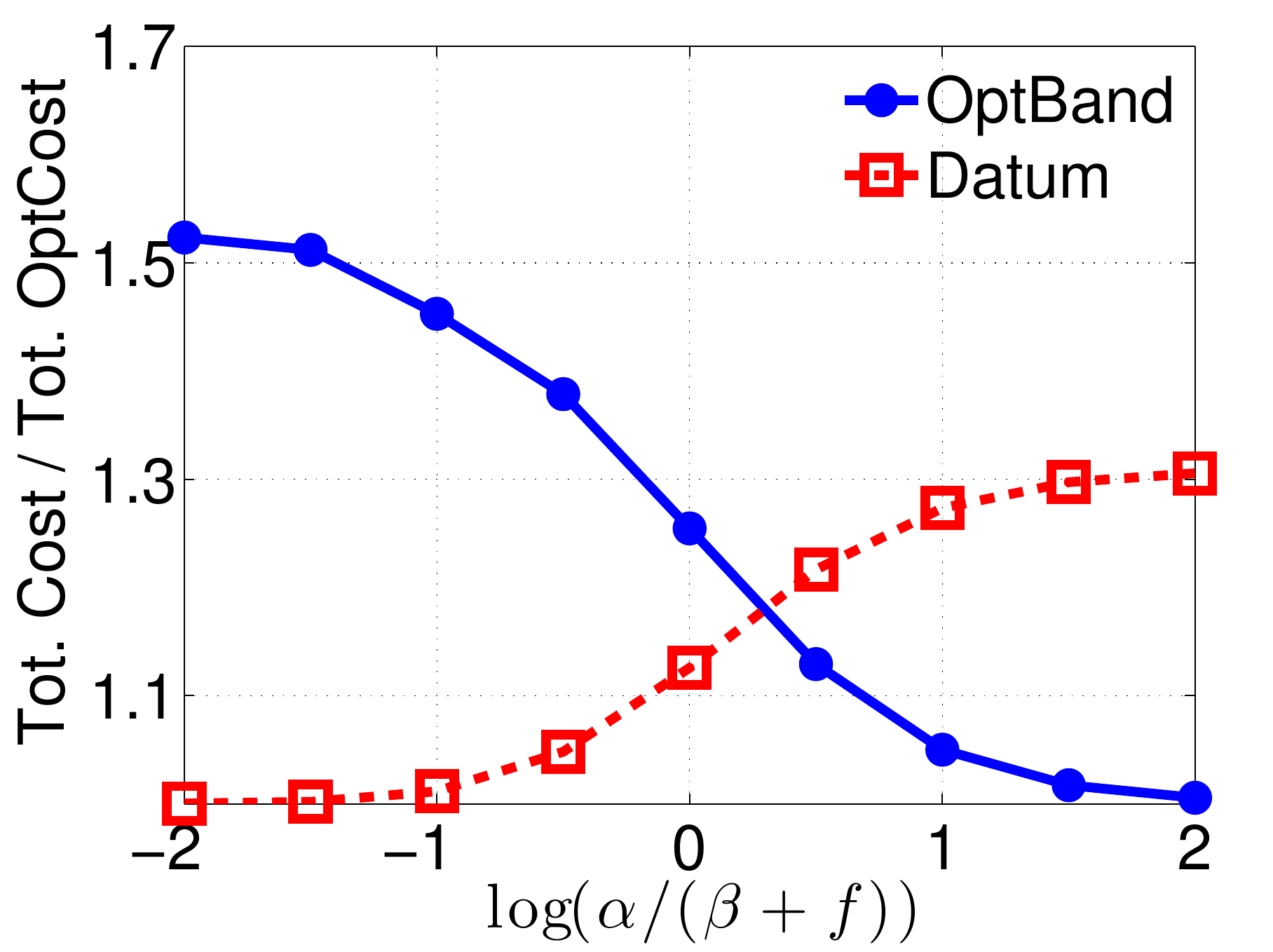}}
\caption{Illustration of the impact of bandwidth and purchasing fees on \nameNospace's performance. NearestDC is excluded because its costs are off-scale. (a) varies the ratio of bandwidth costs (summarized by $\alpha+\beta$) to purchasing costs (summarized by $f$).  (b) varies the ratio of costs internal to the data cloud ($\alpha$) to costs external to the data cloud ($\beta+f$).  Note that in Figure \ref{fig:complexity-plots} the ratios are set to $\log(\frac{\alpha+\beta}{f}) = -0.5$ and $\log(\frac{\alpha}{\beta + f}) = -1.$}
\label{fig:geo-plots}
\end{figure*}

\vspace{2pt}\noindent\textbf{Data purchasing vs.\ bandwidth costs.}  The most important determinant of the magnitude of \nameNospace's cost savings is the relative importance of data purchasing costs. In one extreme, if data is free, then the data purchasing decisions disappear and the problem is simply to do data placement in a manner that minimizes bandwidth costs.  In the other extreme, if data purchasing costs dominate then data placement is unimportant.
In Figure~\ref{fig:geo-plots} we only compare total costs among OptCost, OptBand, and \nameNospace. NearestDC is far worse (more than $5$ times worse than OptCost in some cases) and thus is dropped from the plots. Figure \ref{fig:geo-plots}(a) studies the impact of the relative size of data purchasing and bandwidth costs. When the x-axis is 0, the data purchasing and bandwidth costs of the data center are balanced.  Positive values mean that bandwidth costs dominate and negative values mean that data purchasing costs dominate.  As expected, \nameNospace's cost savings are most dramatic in regimes where data purchasing costs dominate.  Cost savings can be $54\%$ in extreme settings. 
Data purchasing costs are expected to dominate in the future -- for some systems this is already true today.
However, it is worth noting that, in settings where bandwidth costs dominates, \name can deviate from the optimal cost by $10-20\%$ in extreme circumstances, and can be outperformed by the MinBand benchmark.  Of course, \name is not designed for such settings given its prioritization of the minimization of data purchasing costs.

\vspace{2pt}\noindent\textbf{Internal vs.\ external costs.} An important aspect of the design of \name is the decomposition of data purchasing decisions from data placement decisions.  This provides a separation between the internal and external operations (and costs) of \nameNospace.  Given this separation, it is important to evaluate the sensitivity of \nameNospace's design to the relative size of internal and external costs.  

Given that \name prioritizes the optimization of external costs (optimizing them in Step 1, see~\xref{SUBSEC:HEURISTIC}), it is natural to expect that \name performs best when these costs dominate.  This is indeed the case, as illustrated in Figure \ref{fig:geo-plots}(b).  Like in Figure \ref{fig:geo-plots}(a), when the x-axis is 0, the internal and external costs are balanced.  Positive values indicate the internal costs dominate and negative values indicate the external costs dominate. In settings where external costs dominate \name can provide $50\%$ cost savings and be within a few percent of the optimal.  However, in cases when internal costs dominate \name can deviate from the optimal cost by $10-30\%$ in extreme circumstances, and can be outperformed by the MinBand benchmark.  Note that, as data purchasing costs grow in importance, external costs will dominate, and so we can expect that \name will provide near optimal performance in practical settings.

\section{Related work}

Our work focuses on the joint design of data purchasing and data placement  in a geo-distributed cloud data market.  As such, it is related both to recent work on data pricing and to geo-distributed data analytics systems.  Further, the algorithmic problem at the core of our design is the facility location problem, and so our work builds on that literature. We discuss related work in these three areas in the following. 

\vspace{2pt}\noindent\textbf{Data pricing:} The design of data markets has begun to attract increasing interest in recent years, especially in the database community, see~\cite{balazinska2011data} for an overview. The current literature mainly focuses on query-based pricing mechanism designs~\cite{koutris2012query, koutris2013toward, li2014theory} and seldom considers the operating cost of the market service providers (i.e., the data cloud). There is also a growing body of work related to data pricing with differentiated qualities~\cite{fleischer2012approximately, li2014theory, cummings2015accuracy}, often motivated by privacy. See~\xref{subsec:price} for more discussion. This work relates to data pricing on the data provider side and is orthogonal to our discussion in this paper.

\vspace{2pt}\noindent\textbf{Geo-distributed data analytics systems:} As cloud servers are increasingly located in geo-distributed systems, analysis and optimization of data stored in geographically distributed data centers has received increasing attention~\cite{vulimiri2015global, vulimiri2015wanalytics, pu2015low, hung2015scheduling}. Bandwidth constraints~\cite{vulimiri2015global, vulimiri2015wanalytics} as well as latency~\cite{pu2015low} are the two main challenges for system design, and a number of system designs have been proposed, e.g., see~\xref{subsec:data_placement} for more discussion.  Our work builds on the model of geo-distributed data analytics systems in~\cite{pu2015low, vulimiri2015global}, but is distinct from this literature because none of the work on geo-distributed data analytics systems considers the costs associated with purchasing data.

\vspace{2pt}\noindent\textbf{Algorithms for facility location:} Our data cloud cost minimization problem can be viewed as a variant of the uncapacitated facility location problem. Though such problems have been widely studied, most of the results, especially algorithms with constant approximation ratios, require the assumption of metric cost parameters~\cite{CGTS99, GK99, JKV01}, which is not the case in our problem. In contrast, for the non-metric facility location problem the best known algorithm is a greedy algorithm proposed in~\cite{Krarup83}.   Beyond this algorithm, a variety of heuristics have been proposed, however none of the heuristics are appealing for our problem because it is desirable to separate (external) data purchasing decisions from (internal) data placement/replication decisions as much as possible.  As a result we propose a new algorithm, \nameNospace,  which is both near-optimal in practical settings and provides the desired decomposition.  \name may also be valuable more broadly for facility location problems.

\section{Concluding remarks}

This work sits at the intersection of two recent trends: the emergence of online data marketplaces and the emergence of geo-distributed data analytics systems.  Both have received significant attention in recent years across academia and industry, changing the way data is bought and sold and changing how companies like Facebook run queries across geo-distributed databases.  In this paper we study the engineering challenges that come when online data marketplaces are run on top of a geo-distributed data analytics infrastructure.  Such cloud data markets have the potential to be a significant disruptor (as we highlight in \xref{sec:opps}).  However, there are many unanswered economic and engineering questions about their design.  While there has been significant prior work on economic questions, such as how to price data, the engineering questions have been neglected to this point.

In this paper, we have presented the design of a geo-distributed cloud data market: \nameNospace. \name jointly optimizes data purchasing decisions with data placement decisions in order to minimize the overall cost.  While the overall cost minimization problem is NP-hard (via a reduction to/from the facility location problem), \name provides near-optimal performance (within $1.6\%$ of optimal) in realistic settings via a polynomial-time algorithm that is provably optimal in the case of a data cloud running on a single data center.  Additionally, \name provides $> 45\%$ improvement over current design proposals for geo-distributed data analytics systems.  \name works by decomposing the total cost minimization problem into subproblems that allow optimization of data purchasing and data placement separately, which provides a practical route for implementation in real systems.  Further, \name provides a unified solution across systems using per-query pricing or bulk pricing, systems with data replication constraints and/or regulatory constraints on data placement, and systems with SLA constraints on delivery.

\appendix

\section{Proof of Theorem~\ref{T:REDUCTION}}
\label{app:reduction}
To prove Theorem~\ref{T:REDUCTION}, we show a connection between the data cloud cost minimization problem in~\eqref{eqn: ILP} and the uncapacitated facility location problem.  In particular, we show both that the facility location problem can be reduced to a data cloud optimization problem and vice versa.

First, we show that every instance of the uncapacitated facility location problem can be viewed as an instance of~\eqref{eqn: ILP}.

%

Take any instance of the uncapacitated facility location problem (UFLP). Let $I$ be the set of customers, $J$ the set of locations, $\alpha_{ij}$ the cost of assigning customer $i$ to location $j$, and $\beta_j$ the cost of opening facility $j$. Binary variables $y_j = 1$ if and only if facility is open at site $j$, and $x_{j, i} = 1$ if and only if customer $i$ is assigned to location $j$. Then the UFLP can be formulated as following.
\begin{align}\label{eqn: ufl}
\underset{x, y} {\text{min}} \quad &
\sum_{j \in F} \beta_{j}y_{j} +
\sum_{i \in I,j \in F} \alpha_{ij} x_{j,i} \\
 \text{subject to}\nonumber\\
&x_{j,i} \le y_{j},\;\forall i,j\nonumber\\
&\sum_{j \in F} x_{j,i}= 1, \;\forall c\nonumber\\
&x_{j,i}, y_{j} \in \{0, 1\}, \forall i, j\nonumber
\end{align}

Mapping $j$ to $d$ and $i$ to $c$ yields an instance of~\eqref{eqn: ILP} with $|P| = |L| = 1$, $f(l) = 0$ and $w_c(l) = 0$, in which case constraint~\eqref{ILP: quality} becomes trivial.

Next, we show that every instance of~\eqref{eqn: ILP} can be written as an instance of UFLP.

We start by remarking that~\eqref{eqn: ILP} (with $p$ dropped) is equivalent to the following ILP. 
\begin{align}
\underset{x, y} {\text{min}} \quad &
\sum_{d,l=1}^{D,L} \beta_{d}(l) y_{d}(l) +
\sum_{d,l,c=1}^{D, L,C } (f(l) + \alpha_{d,c}(l)) x_{d,c}(l) \\
 \text{subject to}\nonumber\\
&  x_{d,c}(l) \le y_{d}(l),\;\forall c, l, d\nonumber\\
&\sum_{d=1}^D\sum_{l=1}^{L} x_{d,c}(l) = 1, \;\forall c\nonumber\\
&x_{d,c}(l), y_{d}(l) \in \{0, 1\}, \forall c, l,d\nonumber
\end{align}
with $\alpha_{d,c}(l)=M$, for $M$ big enough, whenever $l < w_c$. Indeed, in any feasible solution of~\eqref{eqn: ILP}, we necessarily have $x_{d,c}(l)=0$ whenever $l<w_c$, as each client purchases exactly one quality level and this quality level has to be higher than the minimum required level $w_c$; by setting $\alpha_{d,c}(l)$ big enough, we ensure that any optimal solution must have $x_{d,c}(l)=0$ thus must be feasible for~\eqref{eqn: ILP}, and has the same cost as in~\eqref{eqn: ILP}. Now, take $J=[D] \times [L]$ and $I=[C]$, and the problem can be rewritten as
 \begin{align}
\underset{x, y} {\text{min}} \quad &
\sum_{(d,l) \in J} \beta_{d}(l) y_{d}(l) +
\sum_{(d,l) \in J,c \in I} (f(l) + \alpha_{d,c}(l)) x_{d,c}(l) \\
 \text{subject to} \quad
&  x_{d,c}(l) \le y_{d}(l),\;\forall (d,l) \in J, c \in I\nonumber\\
&\sum_{(d,l) \in J} x_{d,c}(l) = 1, \;\forall c \in I\nonumber\\
&x_{d,c}(l), y_{d}(l) \in \{0, 1\}, \forall c \in I, (d,l) \in J\nonumber
\end{align}
which is an UFLP.

\section{Proof of Theorem~\ref{T:SINGLEDC}}\label{app:singledc}

Assume without loss of generality that all clients can be satisfied by the highest quality level, i.e., $w_c \le q(L), \forall c$. 
Define $C_i = \{c: q(i-1) < w_c \le q(i)\}$ (q(0) = 0 by default). Given these assumptions, clients can be grouped into $L$ categories $\{C_1, C_2, \ldots, C_L\}$ based on their minimum quality level. Note that $C_i \cap C_j = \emptyset, \forall i, j$ and $\cup_{i=1}^L C_i = C$. Without loss of generality, assume $C_i \neq \emptyset,\;\forall i$. 

As the clients in the same group $C_i$ all face exactly the same choice of quality levels and minimum quality requirements,
there must always be an optimal solution in which  the data purchasing decisions of any clients within one category are the same. 

Let us denote the number of clients in category $C_i$ by $S_i$. Denote the purchasing decision of category $C_i$ by $\chi_i$, e.g., $\chi_i(l) = x_c(l), \;\forall l, c\in C_i$, similar to the argument in proof of Theorem~\ref{T:REDUCTION}, we can reformulate~\eqref{eqn: ILP_single} as follows. Note the slight abuse of notation, as clients and their associated required quality level are represented by the same letter, $i$, due to clients in category $C_i$ having minimum quality level $i$ by definition.
\begin{subequations}\label{eqn: CIP}
\begin{align}
\text{minimize}\quad& \sum_{l=1}^L\beta(l)y(l) + \sum_{i=1}^L\sum_{l=i}^L S_if(l)\chi_i(l)\tag{\ref{eqn: CIP}}\\
\text{subject to} \quad &\chi_i(l) \le y(l),\;\forall i, l\label{CIP: existence}\\
\label{CIP: x_equality}
&\sum_{l=i}^{L} \chi_i(l) = 1, \;\forall i\\
&\chi_i(l) \ge 0, \forall i, l\\
&y(l) \ge 0, \;\forall l\\
&\chi_i(l), y(l) \in \{0, 1\}, \forall i, l\label{CIP: integer}
\end{align}
\end{subequations}

Consider the linear relaxation of~\eqref{eqn: CIP}, which drops the $0-1$ integer constraint~\eqref{CIP: integer}. For any optimal solution $\{\chi^{r}_i(l), y^{r}(l)\}$ of the linear relaxation we have the following observations.
	\begin{enumerate}
		\item $\chi^r_L(L) = 1$. \label{LP=IP: y_n}
		\begin{proof}
		From~\eqref{CIP: x_equality}, let $i = L$, then $\chi^r_L(L) = 1$. The intuition behind this is that, since $C_L \neq \emptyset$, highest quality data always has to be purchased to provide service for clients in $C(L)$.
		\end{proof}
		\item $y^r(l) = \max_i\{\chi^r_i(l)\}\in[0, 1]$ and $y^r(L) = 1$. \label{LP=IP: y(c)}
		\begin{proof}
		From \eqref{CIP: existence}, the non-negativity of $\{\beta(l)\}$, and the optimality of $\{y^r(l)\}$, $y^r(l) = \max_i\{\chi^r_i(l)\}$. From the non-negativity of $\{\chi^r_i(l)\}$, $y^r(l) =\max_i\{\chi^r_i(l)\} \le \sum_{l=i}^L \chi^r_i(l)=1$, and $y^r(L) = \chi^r_L(L) = 1$
		\end{proof}
		\item $\forall l \ge i$, if $\sum_{l=i}^L y^r(l) \le 1, \chi^r_i(l) = y^r(l)$; otherwise, $\chi_i(l) = \max\{1-\sum_{k=i}^{l-1}y^r(k), 0\}$.  \label{LP=IP: x_assignment}
		\begin{proof}
		For some fixed $i$, $\{S_i f(l)\}$ is a positive, strictly increasing sequence as $l$ increases. From constraint~\eqref{CIP: existence} and ~\eqref{CIP: x_equality}, $\chi^r_i(l)\le y^r(l)$, and $\sum_{l=i}^L \chi^r_i(l) = 1$. Since $\{\chi^{r}_i(l), y^{r}(l)\}$ is optimal, $\forall l \ge i$, if $\sum_{k=i}^l y^r(k) \le 1, \chi^r_i(l) = y^r(l)$; otherwise, $\chi^r_i(l) = \max\{1-\sum_{k=i}^{l-1}y^r(k), 0\}$.
		\end{proof}
	\end{enumerate}	

Next, define $m_i\in \{i, \ldots, n\}$ such that $\sum_{l=i}^{m_i-1} y^r(l) < 1$, and $\sum_{l=i}^{m_i} y^r(l) \ge 1$. Such an $m_i$ must exist since $y^r(l) \ge 0$ for all $l$  and $y^r(L) = 1$. Recall $\chi^r_L(L) = y^r(L) = 1$. For for any $i = 1, 2, \ldots, L-1$, if the values of $\{y^r(l)\}$ are given, the optimal $\{\chi_i^r(l)\}$ satisfy the following closed form expression:
	\begin{eqnarray}\label{z'_i(l), l_i(l)}
		\chi^r_i(l) =
		\begin{cases}
			y^r(l), & i \le l < m_i\\
			1-\sum_{k=i}^{m_i-1}y^r(k), &l = m_i\\
			0, &m_i < l \le n.
		\end{cases}
	\end{eqnarray}


Note that, if $y^r$ are binary, then $\chi^r$ are binary.
Suppose there exists an optimal solution $\{\chi^r,y^r\}$ with $y^r \not\in \{0, 1\}^L$, in the following we show that there exists a feasible binary solution $\{\chi^*,y^*\}$ of~\eqref{eqn: CIP} such that the objective value generated by $\{\chi^*,y^*\}$ is better than  or equal to that of $\{\chi^r,y^r\}$.

Suppose fractional solution $y^r$ is an optimal solution of the linear relaxation and calculate $m_i$ as in~\eqref{z'_i(l), l_i(l)}. Write $\chi$ as a function of $y$, $\forall i, l$.
	\begin{eqnarray}\label{z_i(l), l_i(l)}
		\chi_i(l) =
		\begin{cases}
			y(l), & i \le l < m_i\\
			1-\sum_{k=i}^{m_i-1}y(k), &l = m_i\\
			0, &m_i < l \le n
		\end{cases}
	\end{eqnarray}
	  Substituting 
\eqref{z_i(l), l_i(l)} in the objective function~\eqref{eqn: CIP}, the objective function becomes a linear combination of $\{y(l)\}$ that we denote $L(y)$.
	
%
Consider the optimization problem in which $\{\chi_i(l)\}$ is expressed as a function of $\{y(l)\}$ in the linear relaxation:
	\begin{subequations}\label{eqn: P}
	\begin{align}
	\text{minimize}\quad& L(y)\tag{\ref{eqn: P}}\\
	\text{subject to}\quad 
	&\sum_{l=i}^{m'_i-1}y(l) \le 1,\;\forall i=1, \ldots, L-1\nonumber\\
	&\sum_{l=i}^{m'_i}y(l) \ge 1,\;\forall i=1, \ldots, L-1\nonumber\\
	&y(l) \ge 0, \;\forall l=1, \ldots, L\nonumber\\
	&y(L) = 1\nonumber
	\end{align}
	\end{subequations}
	The following claims hold:
	\begin{enumerate}
		\item \eqref{eqn: P} is feasible and bounded, and always has an optimal solution at an extreme point.
		\begin{proof}		
		Clearly, $\forall l, y(l) \in [0, 1]$. And starting from $y(L)$, it is easy to construct a feasible solution of~\eqref{eqn: P}. Thus, \eqref{eqn: P} is feasible and bounded, and always has an optimal solution at an extreme point.
		\end{proof}
		\item $\{y^r(l)\}$
 is a feasible solution of~\eqref{eqn: P}.
		\begin{proof}
		Since $\{y^r(l)\}$ is feasible for~\eqref{eqn: CIP}, $y^r(l) \ge 0, \;\forall l$, and $y^r(L) = 1$. By definition of $m_i$, $\sum_{l=i}^{m_i-1}y^r(l) \le 1$, $\sum_{l=i}^{m_i}y^r(l) \ge 1$.
		\end{proof}
		\item Any extreme point $\{y(l)\}$ of~\eqref{eqn: P} is binary.
		\begin{proof}
		Since $y(L) = 1$, we can drop $y(L)$, and write~\eqref{eqn: P} in the following standard linear programming form:
		\begin{align}
		\min_y \quad& L(y) \label{TUM_simple} \\
		\text{s.t.}\quad
		&Ay \le b\nonumber\\
		&y \ge 0\nonumber
		\end{align}
		Note that all entries of $A$ are $0, \pm 1$, and all rows of $A$ has either consecutive $1$s or consecutive $-1$s.  Thus, from \cite{schrijver1998theory}, $A$ is a totally unimodular matrix thus the extreme points of~\eqref{TUM_simple} are all integral. In particular, since all $y(l) \in [0, 1]$, the extreme points of~\eqref{TUM_simple} are all binary.
	\end{proof}
		\item The $\{\chi^*_i(l)\}$ obtained through~\eqref{z_i(l), l_i(l)} corresponding to an optimal binary solution $\{y^*\}$ is also binary.
		\begin{proof}
		 Follows immediately from~\eqref{z_i(l), l_i(l)} and integrality of $\{y^*(l)\}$.
		\end{proof}
		\item $\{\chi^*_i(l), y^*(l)\}$ is a feasible solution of the linear relaxation of~\eqref{eqn: CIP}.
		\begin{proof}
			Follows from~\eqref{z_i(l), l_i(l)} and $\sum_{l=i}^L \chi^*_i(l) = 1$
		\end{proof}

$\{\chi^r_i(l), y^r(l)\}$ and any optimal extreme point $\{\chi^*_i(l), y^*(l)\}$ see their corresponding objective values unchanged between~\eqref{eqn: P} and the relaxation of~\eqref{eqn: CIP} by construction of the $\chi_i(l)$'s. And any such extremal and optimal $\{\chi^*_i(l), y^*(l)\}$ has a better or equal objective value compared to $\{\chi^r_i(l), y^r(l)\}$ in relaxed~\eqref{eqn: CIP}. 
Since $\{\chi^r_i(l), y^r(l)\}$ is optimal for~\eqref{eqn: P}, it implies any optimal extreme point of relaxed~\eqref{eqn: CIP} yields a binary and optimal solution for~\eqref{eqn: P}. This provides a polynomial time algorithm to find such a binary optimal solution, which can be summarized as in~\xref{SUBSEC:HEURISTIC}.
	\end{enumerate}

\section{Proof of Step 2 in~\xref{SUBSEC:HEURISTIC}} \label{app:heu}

In this section we derive the closed form solutions of~\eqref{eqn: closed_form} for the optimization in \eqref{eqn: heu_S2}.  We start by discussing the form of $x_{v,c}(l)$.  Consider the following two cases based on the value of $Y(l)$.
\begin{enumerate}
\item For any quality level $l'$, if $Y(l') = 0$, then $\forall v, \sum\limits_{v=1}^V y_{v}(l') = Y(l') = 0$. From the non-negativity of $y_v(l')$, $\forall v, y_v(l') = 0$. Further, $\forall v, c, x_{v, c}(l') = 0$ from~\eqref{heu_S2: x_less_than_y}.
\item For any quality level $l'$, if $Y(l') = 1$, then from the definition of $y_v(l)$ and $Y(l)$, $\exists ! v'\in V$, such that $y_{v'}(l') = Y(l') = 1$. Recall that $C(l') = \{c: X_c(l') = 1\}$ represents the set of clients that are assigned data with quality level $l'$ by Step 1 in~\xref{SUBSEC:HEURISTIC}.
	\begin{enumerate}
	\item For client $c'\in C(l')$, $X_{c'}(l') = 1$. Since $v'$ is the unique data center set across $V$ such that $y_{v'}(l') = 1$, from~\eqref{heu_S2: x_less_than_y} and~\eqref{heu_S2: x_equal_1}, $x_{v', c'}({l'}) = 1$ and $x_{v, c'}({l}) = 0, \;\forall v\neq v'$ or~$l\neq l'$. In other words, $x_{v, c'}({l'}) = y_v(l'), \,\forall v\in V, c\in C(l').$
	\item For client $c\notin C(l')$, $X_{c}(l') = 0$. From the definition of $X_c(l')$, $x_{v, c}(l') = 0, \,\forall v.$
	\end{enumerate}
\end{enumerate}

In all above cases, the optimal solution $\{x_{v,c}(l), y_v(l)\}$ of~\eqref{eqn: heu_S2}  satisfies the following:
\begin{subequations}\label{heu_S2: x_solution}
\begin{flalign}\tag{\ref{heu_S2: x_solution}}
x_{v, c}(l) =
\begin{cases}
y_v(l), &\text{if}~c \in C(l),\\
0, &\text{otherwise.}
\end{cases}
\end{flalign}
\end{subequations}

Next, we use this form for $x_{v,c}(l)$ to derive $y_v(l)$.  After substituting~\eqref{heu_S2: x_solution} into~\eqref{eqn: heu_S2},  most constraints become trivial due to the form of~\eqref{heu_S2: x_solution} and the optimality of $X_c(l)$ and $Y(l)$. And we only need to optimize the objective function with the constraints stating that  $y_v(l)$ is binary, and $\sum_v y_v(l) = Y(l)$. Thus, we only need to optimize the following problem.
\begin{subequations}\label{eqn: heu_S2_final}
\begin{align}
\text{minimize}\quad& \sum_{l: Y(l) = 1} \sum_{v=1}^V \beta_v(l) y_v(l)\nonumber\\ &  + \sum_{l: Y(l) = 1}\sum_{c\in C(l)}\sum_{v=1}^V (\alpha_{v, c}(l)+f(l))y_v(l)\nonumber\\
	\text{subject to}\quad 
&\sum_{v=1}^V y_v(l) = Y(l), \forall v, c, l\nonumber\\
&y_{v}(l) \in \{0, 1\}, \forall v, c, l\nonumber
\end{align}
\end{subequations}
The above optimization can be decoupled by $l$ and optimized across $v$, yielding the following closed form solution.

\begin{subequations}\label{heu_S2: x_solution}
\begin{flalign}\tag{\ref{heu_S2: x_solution}}
y_v(l) =
\begin{cases}
1, &\text{if}~Y(l) = 1~\text{and}~\\
   & v = \text{argmin}\{\beta_v(l) + \sum_{c\in C(l)} \alpha_{v, c}(l)\},\\
0, &\text{otherwise.}
\end{cases}
\end{flalign}
\end{subequations}

\section{Bulk Data Contracting}\label{app:bulk}
In bulk data contracting, the data cloud only has to pay a one-time fee $f(l, p)$ for data $q(l, p)$, no matter how many times the data is replicated on the cloud and transferred to clients. Compared to per-query contracting, the main difference lies in the purchasing fees modeling. Defining $z(l, p)   \in \{0,1\}$ to be equal to $1$ if and only if data of quality $q(l,p)$ from data provider $p$ is transferred to the data cloud, the whole optimization problem can still be formulated in a form similar to~\eqref{eqn: ILP}, with the purchasing costs now given by~\eqref{eqn: bulk_fee} and with the addition of the following constraint:
\begin{align}
 y_{p, d}(l) \le z(l, p), \;\forall c, l, p, d
\end{align}
This constraint states that any data placed in the data cloud must have been purchased by the data cloud. As in the per-query contracting case,  the data purchasing/placement decision for data from one data provider does not impact the data purchasing/placement decision for any other data providers. Thus, we drop the index $p$ in the following.

In general, the cost minimization problem for bulk contracting is NP-hard.  To be specific, the 1-level UFLP can reduce to the cost minimization problem for a geo-distributed data cloud, and the cost minimization problem can reduce to the 2-level UFLP in the bulk case. In the 2-level UFLP, facilities are organizing on $2$ levels, $J_1 \times J_2$; each customer $i \in I$ has to be assigned to a valid path $p \in J_1 \times J_2$. A pass is valid if and only if both facilities are open along the path. More details on the 2-level UFLP can be found in~\cite{zhang2006approximating}.

The first reduction follows directly from the first part of the proof for Theorem~\ref{T:REDUCTION}. It can be easily proved by defining facilities in $J_1$ to be the quality levels, and using the same reformulation as the second part of the proof for Theorem~\ref{T:REDUCTION} for the facilities in $J_2$, i.e. define facilities in$J_2$ to be pairs of quality levels and data centers. In the reduction, a facility $j_1 \in J_1$ is open if and only if the corresponding quality level $l$ is purchased, and a facility $j_2 \in J_2$ is opened if and only if data of quality level $l$ is placed in data center $d$.

While the cost minimization in bulk contracting is generally hard, it can be solved optimally in  both the single data center and the geo-distributed data cloud settings under certain assumptions. 

For the single data center case, we always have $z(l) = y(l)$ for all quality level $l$ - this follows immediately from dropping the dependence of $y_d(l)$ in $d$, implying that $z(l)$ is only lower-bounded by $y(l)$ in the constraints. Furthermore, if the execution costs are the same across quality levels, the cost minimization problem can be formulated as follows:
\begin{subequations}\label{eqn: ILP_single_bulk}
\begin{align}
\text{minimize}\quad& \sum_{l=1}^L\left(\beta(l) + f(l)\right)y(l) \tag{\ref{eqn: ILP_single_bulk}}\\
\text{subject to} \quad
&x_{c}(l) \le y(l),\;\forall c, l \nonumber \\
&\sum_{l=w_c}^{L} x_{c}(l) = 1, \;\forall c \nonumber \\
&x_{c}(l) \ge 0, \forall c, l \nonumber\\
&y(l) \ge 0, \;\forall l \nonumber\\
&x_{c}(l), y(l) \in \{0, 1\}, \forall c, l \nonumber
\end{align}
\end{subequations}
Since the decisions for variables $\{x_c(l)\}$ do not affect the objective value,~\eqref{eqn: ILP_single_bulk} can be written as follows:
\begin{subequations}\label{eqn: ILP_single_bulk_s}
\begin{align}
\text{minimize}\quad&\sum_{l=1}^L\left(\beta(l) + f(l)\right)y(l) \tag{\ref{eqn: ILP_single_bulk_s}}\\
\text{subject to} \quad
&\sum_{l=w_c}^{L} y(l) \geq 1, \;\forall l, c \nonumber \\ 
& y(l) \in \{0, 1\}, \forall l \nonumber
\end{align}
\end{subequations}

Since there are customers buying the highest quality level, the highest level quality $L$ is always purchased by the data cloud and $y(L) = 1$ in any feasible solution. Since all customers are satisfied and all costs are non-negative, an optimal solution for~\eqref{eqn: ILP_single_bulk_s} is $y(L) = z(L) = 1$, $x_c(L) = 1$ with all other variables are set to $0$. The result implies the data cloud will only purchase the highest quality level of data and serve that data to every customers.  

For a geo-distributed data cloud, the cost minimization problem is generally hard. However, if we assume the operation cost and execution cost are independent of $l$, i.e., $\beta_d(l) = \beta_d$ and $\alpha_{d, c}(l) = \alpha_{d, c}$, it is easy to show that the optimal solution will only purchase the highest quality data as in the single data center case. We can then use Step $2$ in~\xref{SUBSEC:HEURISTIC} to give an optimal solution to the data placement problem.


\begin{thebibliography}{42}
\providecommand{\natexlab}[1]{#1}
\providecommand{\url}[1]{\texttt{#1}}
\expandafter\ifx\csname urlstyle\endcsname\relax
  \providecommand{\doi}[1]{doi: #1}\else
  \providecommand{\doi}{doi: \begingroup \urlstyle{rm}\Url}\fi

\bibitem[Azu(2015)]{Azure}
{Microsoft Azure}.
\newblock \url{https://azure.microsoft.com/en-us/}, 2015.

\bibitem[Fac(2015)]{Factual}
Factual.
\newblock \url{https://www.factual.com/}, 2015.

\bibitem[Inf(2015)]{Infochimps}
Infochimps.
\newblock \url{http://www.infochimps.com/}, 2015.

\bibitem[Xig(2015)]{Xignite}
Xignite.
\newblock \url{http://www.xignite.com/}, 2015.

\bibitem[IUP(2015)]{IUPHAR}
{The IUPHAR/BPS Guide to Pharmacology}.
\newblock \url{http://www.guidetopharmacology.org/}, 2015.

\bibitem[Koutris et~al.(2012{\natexlab{a}})Koutris, Upadhyaya, Balazinska,
  Howe, and Suciu]{koutris2012query}
P.~Koutris, P.~Upadhyaya, M.~Balazinska, B.~Howe, and D.~Suciu.
\newblock {Query-based Data Pricing}.
\newblock In \emph{Proceedings of the 31st symposium on Principles of Database
  Systems}, 2012{\natexlab{a}}.

\bibitem[Koutris et~al.(2013)Koutris, Upadhyaya, Balazinska, Howe, and
  Suciu]{koutris2013toward}
P.~Koutris, P.~Upadhyaya, M.~Balazinska, B.~Howe, and D.~Suciu.
\newblock {Toward Practical Query Pricing with QueryMarket}.
\newblock In \emph{SIGMOD}, 2013.

\bibitem[Fleischer and Lyu(2012)]{fleischer2012approximately}
L.~Fleischer and Y.~Lyu.
\newblock {Approximately Optimal Auctions for Selling Privacy when Costs are
  Correlated with Data}.
\newblock In \emph{Proceedings of the 13th ACM Conference on Electronic
  Commerce}, 2012.

\bibitem[Li et~al.(2014)Li, Li, Miklau, and Suciu]{li2014theory}
C.~Li, D.~Li, G.~Miklau, and D.~Suciu.
\newblock {A Theory of Pricing Private Data}.
\newblock \emph{ACM Transactions on Database Systems}, 2014.

\bibitem[Pu et~al.(2015)Pu, Ananthanarayanan, Bodik, Kandula, Akella, Bahl, and
  Stoica]{pu2015low}
Q.~Pu, G.~Ananthanarayanan, P.~Bodik, S.~Kandula, A.~Akella, P.~Bahl, and
  I.~Stoica.
\newblock {Low Latency Geo-distributed Data Analytics}.
\newblock In \emph{SIGCOMM}, 2015.

\bibitem[Vulimiri et~al.(2015{\natexlab{a}})Vulimiri, Curino, Godfrey,
  Karanasos, and Varghese]{vulimiri2015wanalytics}
A.~Vulimiri, C.~Curino, B.~Godfrey, K.~Karanasos, and G.~Varghese.
\newblock {WANalytics: Analytics for a Geo-distributed Data-intensive World}.
\newblock In \emph{CIDR}, 2015{\natexlab{a}}.

\bibitem[Vulimiri et~al.(2015{\natexlab{b}})Vulimiri, Curino, Godfrey, Padhye,
  and Varghese]{vulimiri2015global}
A.~Vulimiri, C.~Curino, B.~Godfrey, J.~Padhye, and G.~Varghese.
\newblock {Global Analytics in the Face of Bandwidth and Regulatory
  Constraints}.
\newblock In \emph{NSDI}, 2015{\natexlab{b}}.

\bibitem[Krarup and Pruzan(1983)]{Krarup83}
J.~Krarup and P.~Pruzan.
\newblock {The Simple Plant Location Problem: Survey and Synthesis}.
\newblock \emph{{European Journal of Operational Research}}, 1983.

\bibitem[Charikar et~al.(1999)Charikar, Guha, Tardos, and Shmoys]{CGTS99}
M.~Charikar, S.~Guha, \'{E}. Tardos, and D.~Shmoys.
\newblock {A Constant-factor Approximation Algorithm for the K-median Problem
  (Extended Abstract)}.
\newblock In \emph{STOC}, 1999.

\bibitem[Guha and Khuller(1999)]{GK99}
S.~Guha and S.~Khuller.
\newblock {Greedy Strikes Back: Improved Facility Location Algorithms}.
\newblock \emph{Journal of Algorithms}, 1999.

\bibitem[Jain and Vazirani(2001)]{JKV01}
K.~Jain and V.~Vazirani.
\newblock {Approximation Algorithms for Metric Facility Location and k-Median
  Problems Using the Primal-dual Schema and Lagrangian Relaxation}.
\newblock \emph{J. ACM}, 2001.

\bibitem[Hochbaum(1982)]{Hochbaum82}
D.~Hochbaum.
\newblock {Heuristics for the Fixed Cost Median Problem}.
\newblock \emph{Math. Program.}, 1982.

\bibitem[Vazirani(2001)]{Vazirani2001}
V.~Vazirani.
\newblock \emph{{Approximation Algorithms}}.
\newblock Springer, 2001.

\bibitem[Uriel(1998)]{Feige98}
F.~Uriel.
\newblock {A Threshold of ln \emph{n} for Approximating Set Cover}.
\newblock \emph{J. {ACM}}, 1998.

\bibitem[Erlenkotter(1978)]{Erlenkotter78}
D.~Erlenkotter.
\newblock {A Dual-Based Procedure for Uncapacitated Facility Location}.
\newblock \emph{Operations Research}, 1978.

\bibitem[Beasley(1993)]{Beasley93}
J.~Beasley.
\newblock {Lagrangean Heuristics for Location Problems}.
\newblock \emph{European Journal of Operational Research}, 1993.

\bibitem[Al-Sultan and Al-Fawzan(1999)]{SF99}
K.~Al-Sultan and M.~Al-Fawzan.
\newblock {A Tabu Search Approach to the Uncapacitated Facility Location
  Problem}.
\newblock \emph{{Annals of Operations Research}}, 1999.

\bibitem[Korkel(1989)]{Korkel89}
M.~Korkel.
\newblock {On the Exact Solution of Large-scale Simple Plant Location Problems
  }.
\newblock \emph{{European Journal of Operational Research }}, 1989.

\bibitem[Tuzun and Burke(1999)]{Tuzun99}
D.~Tuzun and L.~Burke.
\newblock {A Two-phase Tabu Search Approach to the Location Routing Problem }.
\newblock \emph{{European Journal of Operational Research }}, 1999.

\bibitem[Ghosh(2003)]{Ghosh03}
D.~Ghosh.
\newblock {Neighborhood Search Heuristics for the Uncapacitated Facility
  Location Problem }.
\newblock \emph{{European Journal of Operational Research }}, 2003.

\bibitem[Koutris et~al.(2012{\natexlab{b}})Koutris, Upadhyaya, Balazinska,
  Howe, and Suciu]{koutris2012demonstration}
P.~Koutris, P.~Upadhyaya, M.~Balazinska, B.~Howe, and D.~Suciu.
\newblock {QueryMarket Demonstration: Pricing for Online Data Markets}.
\newblock \emph{Proceedings of the VLDB Endowment}, 2012{\natexlab{b}}.

\bibitem[Vis(2015)]{Visipedia}
{Visipedia Project}.
\newblock \url{http://www.vision.caltech.edu/visipedia/}, 2015.

\bibitem[Balazinska et~al.(2013)Balazinska, Howe, Koutris, Suciu, and
  Upadhyaya]{balazinska2013discussion}
M.~Balazinska, B.~Howe, P.~Koutris, D.~Suciu, and P.~Upadhyaya.
\newblock A discussion on pricing relational data.
\newblock In \emph{{In Search of Elegance in the Theory and Practice of
  Computation}}. 2013.

\bibitem[Balazinska et~al.(2011)Balazinska, Howe, and
  Suciu]{balazinska2011data}
M.~Balazinska, B.~Howe, and D.~Suciu.
\newblock {Data Markets in the Cloud: An Opportunity for the Database
  Community}.
\newblock \emph{Proceedings of the VLDB Endowment}, 2011.

\bibitem[Cummings et~al.(2015)Cummings, Ligett, Roth, Wu, and
  Ziani]{cummings2015accuracy}
R.~Cummings, K.~Ligett, A.~Roth, Z.~Wu, and J.~Ziani.
\newblock {Accuracy for Sale: Aggregating Data with a Variance Constraint}.
\newblock In \emph{ITCS}, 2015.

\bibitem[Corbett et~al.(2013)Corbett, Dean, Epstein, Fikes, Frost, Furman,
  Ghemawat, Gubarev, Heiser, Hochschild, et~al.]{corbett2013spanner}
J.~Corbett, J.~Dean, M.~Epstein, A.~Fikes, C.~Frost, J.~Furman, S.~Ghemawat,
  A.~Gubarev, C.~Heiser, P.~Hochschild, et~al.
\newblock {Spanner: Google's Globally Distributed Database}.
\newblock \emph{ACM Transactions on Computer Systems}, 2013.

\bibitem[Gupta et~al.(2014)Gupta, Yang, Govig, Kirsch, Chan, Lai, Wu, Dhoot,
  Kumar, Agiwal, et~al.]{gupta2014mesa}
A.~Gupta, F.~Yang, J.~Govig, A.~Kirsch, K.~Chan, K.~Lai, S.~Wu, S.~Dhoot,
  A.~Kumar, A.~Agiwal, et~al.
\newblock {Mesa: Geo-replicated, Near Real-time, Scalable Data Warehousing}.
\newblock \emph{Proceedings of the VLDB Endowment}, 2014.

\bibitem[Rabkin et~al.(2014)Rabkin, Arye, Sen, Pai, and
  Freedman]{rabkin2014aggregation}
A.~Rabkin, M.~Arye, S.~Sen, V.~Pai, and M.~Freedman.
\newblock {Aggregation and Degradation in JetStream: Streaming Analytics in the
  Wide Area}.
\newblock In \emph{NSDI}, 2014.

\bibitem[Dwork(2011)]{dwork2011differential}
C.~Dwork.
\newblock {Differential Privacy}.
\newblock In \emph{Encyclopedia of Cryptography and Security}. 2011.

\bibitem[Wiener and Boston(2014)]{facebook2014}
J.~Wiener and N.~Boston.
\newblock {Facebook's top open data problems.}
\newblock \url{https://research. facebook.com/blog/1522692927972019/
  facebook-s-top-open-data-problems/}, 2014.

\bibitem[Lee et~al.(2012)Lee, Lin, Liu, Lorek, and Ryaboy]{lee2012unified}
G.~Lee, J.~Lin, C.~Liu, A.~Lorek, and D.~Ryaboy.
\newblock {The Unified Logging Infrastructure for Data Analytics at Twitter}.
\newblock \emph{Proceedings of the VLDB Endowment}, 2012.

\bibitem[Bertsimas and Tsitsiklis(1997)]{bertsimas1997introduction}
D.~Bertsimas and J.~Tsitsiklis.
\newblock \emph{{Introduction to Linear Optimization}}.
\newblock 1997.

\bibitem[Dat(2012)]{Datacenterknowledge}
{Google Data Center FAQ}.
\newblock
  \url{http://www.datacenterknowledge.com/archives/2012/05/15/google-data-center-faq/},
  2012.

\bibitem[Newman(2005)]{newman2005power}
M.~Newman.
\newblock {Power Laws, Pareto Distributions and Zipf's Law}.
\newblock \emph{Contemporary physics}, 2005.

\bibitem[Hung et~al.(2015)Hung, Golubchik, and Yu]{hung2015scheduling}
C.~Hung, L.~Golubchik, and M.~Yu.
\newblock {Scheduling Jobs across Geo-distributed Datacenters}.
\newblock In \emph{Proceedings of the 6th ACM Symposium on Cloud Computing},
  2015.

\bibitem[Schrijver(1998)]{schrijver1998theory}
A.~Schrijver.
\newblock \emph{{Theory of Linear and Integer Programming}}.
\newblock John Wiley \& Sons, 1998.

\bibitem[Zhang(2006)]{zhang2006approximating}
J.~Zhang.
\newblock {Approximating the two-level facility location problem via a
  quasi-greedy approach}.
\newblock \emph{Mathematical Programming}, 2006.

\end{thebibliography}


\end{document}